\documentclass{article}
\usepackage{jheppub}
\usepackage{physics}
\usepackage{tikz}
\usepackage{mleftright}

\newcommand\titlemath[1]{\texorpdfstring{#1}{Lg}}

\begin{document}

\title{On type II string theory on \titlemath{$AdS_3\times S^3\times T^4$} and symmetric orbifolds}

\author[a,b]{Ofer Aharony}
\emailAdd{ofer.aharony@weizmann.ac.il}
\author[a]{and Erez Y.~Urbach}
\emailAdd{erez.urbach@gmail.com}
\affiliation[a]{Department of Particle Physics and Astrophysics,
Weizmann Institute of Science, Rehovot 7610001, Israel}
\affiliation[b]{School of Natural Sciences, Institute for Advanced Study, Princeton 08540, NJ, USA}
\abstract{We discuss in detail the $1+1$-dimensional superconformal field theory dual to type II string theory on $AdS_3\times S^3\times T^4$, emphasizing the string theoretic aspects of this duality. For one unit of NS-NS 5-brane flux ($Q_5=1$), this string theory has been suggested to be dual to a grand-canonical ensemble of $T^{4N}/S_N$ free symmetric orbifold CFTs.
We show how the string genus expansion emerges to all orders for the free orbifold grand-canonical correlation functions. We also discuss how the strong coupling limit of the NS-NS string theory arises (even at large $N$) in the free orbifold description, and argue why this limit does not have a weakly coupled R-R description. 
The dual CFT includes (for all values of $Q_5$) an extra $T^4$ factor that is decoupled from perturbative string theory. We discuss the exactly marginal deformations that relate the different values of $Q_5$, including the precise $J{\bar J}$ deformations mixing this extra $T^4$ with the symmetric orbifold.
}

\maketitle

\section{Introduction and summary}\label{sec:intro}

The duality between type II string theory on $AdS_3\times S^3\times T^4$ (which can be viewed as the near-horizon limit of $Q_1$ D1-branes and $Q_5$ D5-branes wrapped on $T^4$) and a specific $1+1$-dimensional ${\cal N}=(4,4)$ superconformal field theory (SCFT) (the ``D1-D5 SCFT") is one of the original examples of the AdS/CFT correspondence \cite{Maldacena:1997re,Aharony:1999ti}. It shares many features with other examples of the AdS/CFT correspondence; for $Q_1 \gg Q_5 \gg 1$ the theory has one limit of its continuous parameters where it is described by string theory on a weakly coupled and weakly curved background, and another limit where it is a free field theory (a free symmetric orbifold $(T^4)^{Q_1 Q_5} / S_{Q_1 Q_5}$). However, it also has several advantages compared to higher-dimensional examples, most notably the fact that when the background has purely Neveu-Schwarz (NS-NS) charges the string worldsheet theory is under very good control, either in the Ramond-Neveu-Schwarz (RNS) formalism \cite{Maldacena:1998bw,Evans:1998qu,Giveon:1998ns,deBoer:1998gyt,Kutasov:1999xu,Giveon:2001up,Cho:2018nfn,Giribet:2018ada} (for $Q_5 > 1$) or in the hybrid formalism \cite{Berkovits:1999im,Gaberdiel:2011vf,Eberhardt:2018ouy,Gaberdiel:2021njm}.

Many features of this theory, such as its moduli space \cite{Larsen:1999uk}, chiral ring \cite{Dabholkar:2007ey,deBoer:2008ss}, protected \cite{Lunin:2001pw, Pakman:2007hn, Pakman:2009ab, Baggio:2012rr,Gaberdiel:2022oeu,Martinec:2022okx,Iguri:2023khc} and unprotected \cite{Giribet:2001ft,Maldacena:2001km,Troost:2002wk, Gaberdiel:2007vu,Giribet:2007wp,Taylor:2007hs,Cardona:2009hk,Cardona:2010qf,Dei:2021xgh,Dei:2021yom,Dei:2022pkr,Dei:2023ivl} correlators, and its perturbative string spectrum \cite{Maldacena:2000hw, Gaberdiel:2021njm} were studied and understood over the years. However, some of its features, in particular features related to string perturbation theory, are more subtle. In this paper we review what is known about this theory, highlighting three specific confusing issues and how they are resolved.

We begin in section \ref{sec:review} by reviewing this theory from various different points of view -- the supergravity approximation, the CFT, and the string worldsheet. In particular, we describe the parameter space of the theory and the relations between theories with different (relatively prime) values of $Q_1$ and $Q_5$ (and the same $N = Q_1 Q_5$).

In section \ref{sec:st_grand_can} we review one of the strange features of this theory, the fact that perturbative string theory in the background with only NS-NS fluxes on $AdS_3$ and on $S^3$ is not dual to a specific SCFT but rather to a grand-canonical ensemble of SCFTs (with different values of $Q_1$). This fact was suspected based on the worldsheet in the RNS formalism for $Q_5 > 1$ \cite{Kutasov:1999xu,Kim:2015gak}. For $Q_5=1$, where the theory is dual to a free symmetric orbifold, the relation between string theory and the SCFT can be analyzed in great detail. We describe (following \cite{Lunin:2000yv,Pakman:2009zz,Pakman:2009mi}; see also \cite{Pakman:2009ab,Giribet:2007wp,Eberhardt:2019ywk}) how correlation functions of free symmetric orbifolds ${\rm Sym}_N({\cal M})$ have a $1/N$ expansion. However, this expansion does not map directly to a string theory, while its grand-canonical version does, and we write in detail the relation between string theory and CFT correlation functions. The main result is equation \eqref{eq:grand_Z}, which shows how appropriately defined grand-canonical correlation functions exhibit an exact string genus expansion (generalizing the relation between the partition functions, discussed in \cite{Dijkgraaf:1996xw,Bantay:2000eq,Eberhardt:2021jvj}). We comment on the generalization of this relation to $Q_5 > 1$.

In section \ref{sec:breakdown} we discuss the fact that even though the $Q_5=1$ background (with vanishing Ramond-Ramond (R-R) scalars) maps to a free orbifold theory for any value of its parameters, with the $1/N$ expansion of the orbifold mapping to perturbative string theory in the NS-NS description of this background, there is a region in parameter space where the NS-NS string becomes strongly coupled and an S-dual string seems to become weakly coupled. We show that in this region of parameter space the $1/N$ expansion of the free orbifold breaks down, and discuss the fact that even though naively the string coupling in the dual R-R description can become arbitrarily weakly coupled, this description is never really weakly coupled (similar to the behavior of type IIB string theory on $AdS_5\times S^5$ with a small integer flux $N$, which does not become weakly coupled even when $g_s \to 0$). 

Finally, in section \ref{sec:decoupled_sec}, we describe a mysterious feature of the CFT, which is that it includes an extra ${\hat T}^4$ factor, which is completely decoupled when the rest of the CFT is a free symmetric orbifold
(for general values of the parameters, this factor only couples to the rest of the SCFT through $J{\bar J}$ deformations). Usually, string theory does not allow for decoupled sectors (since all states couple to gravity), but in this case, the decoupled sector is a topological Chern-Simons theory on $AdS_3$, and we discuss to what extent and how it couples to the rest of string theory for general values of the parameters.

Our analysis suggests various interesting future directions. When $N$ is not prime, the CFT has a singular limit corresponding to every $N = Q_1 Q_5$ factorization with $Q_5 > 1$, in which its spectrum becomes continuous (described by ``long strings'' on $AdS_3$ \cite{Seiberg:1999xz}). The CFT also has another singular limit where it is described by the orbifold $(T^4)^N/S_N$ with a vanishing theta angle at the $\mathbb{Z}_2$ singularity, and it would be nice to know if this limit has some controllable string theory description. Our results in section \ref{sec:decoupled_sec} suggest various properties of string theory on $AdS_3$ and its D-branes, and it would be nice to confirm these properties directly. It would also be interesting to understand in more detail the precise grand-canonical ensemble that is dual to string theory on the NS-NS background with $Q_5>1$.

The discussion of the free orbifold limit can be generalized to other purely NS-NS type IIB backgrounds with $1$ unit of 3-form flux on an $S^3$, such as $AdS_3\times S^3\times K3$ or $AdS_3\times S^3\times S^3\times S^1$ \cite{Eberhardt:2019niq}. However, in those cases there is no direct relation to other values of the flux, so it is not clear what can be said about them. It would be interesting to understand what can be said about other backgrounds, like orbifolds of $AdS_3$ and/or $S^3$ \cite{Kutasov:1998zh,Martinec:2001cf,Martinec:2023zha,Gaberdiel:2023dxt}, type IIA string theory on $AdS_3\times S^3\times K3$ \cite{Hohenegger:2008du} and heterotic string theories (including non-supersymmetric ones \cite{Baykara:2022cwj,Fraiman:2023cpa}) on $AdS_3$. 

Last but not least, it would be nice if the detailed understanding of the duality between free orbifolds and string theory could be generalized to the case of free gauge theories with continuous gauge groups.

\section{A review of type II string theory on \titlemath{$AdS_3\times S^3\times T^4$} and its dual CFT}
\label{sec:review}

One way to obtain type II string theory on $AdS_3\times S^3\times T^4$ is by considering the near-horizon limit \cite{Maldacena:1997re} of BPS strings preserving 2d ${\cal N}=(4,4)$ supersymmetry in type II string theory on $T^4$ (coming from fundamental strings, wrapped NS5-branes or 8 types of wrapped D-branes). The type IIA and type IIB cases are related by T-duality so there is no need to discuss them separately, for convenience we will use the type IIB language throughout this paper. Type II string theory on $T^4$ has an $SO(5,5,\mathbb{Z})$ U-duality group, under which the ten string charges transform as a vector. By a U-duality transformation one can go to a frame where we have only D-string charge $Q_1$ and wrapped D5-brane charge $Q_5$, and we will focus on this case in our discussion. By S-duality these configurations are identical to configurations carrying $Q_1$ units of fundamental string charge and $Q_5$ units of wrapped NS5-brane charge. In this section, we review what is known about these configurations from three different points of view -- supergravity, the string worldsheet, and the dual CFT. We will then review how U-duality relates different $AdS_3$ backgrounds.

\subsection{Supergravity solutions}\label{sec:sugra}

The near-horizon limit of $Q_1$ D1-branes and $Q_5$ D5-branes wrapped on a $T^4$ is $AdS_3\times S^3\times T^4$. String theory on $T^4$ has 25 massless scalar fields, including the metric and $B$-field on the $T^4$ (16 scalars), the dilaton (1 scalar), and 8 R-R scalars. When both $Q_1$ and $Q_5$ are non-zero (as we will assume throughout this paper) 5 of these scalars (4 of the R-R scalars and the overall volume of the $T^4$) are fixed in the near-horizon limit, while the other 20 scalars remain as moduli.

We will begin by focusing on the case where the R-R scalars vanish, while the string coupling takes some arbitrary value $g_{10}$, and the torus has some fixed shape and $B$-field. In this case the near-horizon limit is $AdS_3\times S^3\times T^4$, with R-R 3-form flux on $AdS_3$ and on $S^3$, where the radius $R$ of $AdS_3$ and $S^3$ and the volume of $T^4$ are given by (up to numerical constants that will not be important for our discussion) :
\begin{equation}
	\frac{R^2}{\alpha'} = g_{10} Q_5 = g_6 \sqrt{Q_1 Q_5}, \quad \frac{\text{vol}(T^4)}{\alpha'^2} = \frac{Q_1}{Q_5}, \quad 
	g^2_{10} = g_6^2 \frac{Q_1}{Q_5}.
\end{equation}
Here we defined the six-dimensional string coupling $g_6$ in the standard way, and the expressions are derived from the supergravity solution for these branes, so they are reliable when $R$ and $\text{vol}(T^4)$ are much larger than the string scale. Naively, perturbative string theory in this R-R background is reliable when $g_6, g_{10} \ll 1$, while supergravity is valid when $Q_1 \gg Q_5$, $g_{10} Q_5 \gg 1$. The 6- and 10- dimensional Newton constants (in AdS units) in the background described above are
\begin{equation}\label{eq:newtons}
    \frac{G_N^{(6)}}{R^4} = \frac{1}{Q_1 Q_5}, \quad \frac{G_N^{(10)}}{R^8} = \frac{1}{g_6^2 Q_1 Q_5^3}.
\end{equation}

Performing S-duality, we find a purely NS-NS background (that also arises as the near-horizon limit of $Q_1$ fundamental strings and $Q_5$ NS5-branes). In this background, there is NS-NS 3-form flux on $AdS_3$ and on $S^3$ instead of R-R 3-form flux. While the ten-dimensional string coupling in this new frame, $g_{10}'=1/g_{10}$, is still a free parameter, the six-dimensional string coupling is now fixed, while the volume of the torus in string units in this frame is a free parameter which we will denote by $v_4$, related to the R-R frame by $v_4 = g_6^{-2}$. The new background is given by
\begin{equation} \label{eq:ns_ns_desc}
	\frac{R^2}{\alpha'} = Q_5, \quad \frac{\text{vol}(T^4)}{\alpha'^2} = v_4, \quad g_6'^2 = \frac{Q_5}{Q_1}, \quad g_{10}'^2 = v_4 \frac{Q_5}{Q_1}.
\end{equation}
In this background all 16 scalars parameterizing the metric and $B$ field on the $T^4$ are massless moduli, as well as 4 R-R scalars (which we are still setting to zero). Perturbative string theory is now valid whenever $g_6', g_{10}' \ll 1$, while supergravity is valid for $Q_5, v_4 \gg 1$. The NS-NS description is invariant under a T-duality transformation inverting the four cycles of the torus, so we can choose without loss of generality $v_4 \ge 1$ (implying that for fixed $Q_1$, $Q_5$, the ten-dimensional string coupling cannot be arbitrarily small).
Newton's constants remain the same, and in terms of $v_4$ they are given by \eqref{eq:newtons}
\begin{equation}\label{eq:newtons_V}
    \frac{G_N^{(6)}}{R^4} = \frac{1}{Q_1 Q_5}, \quad \frac{G_N^{(10)}}{R^8} = \frac{v_4}{Q_1 Q_5^3}.
\end{equation}

\subsection{The dual conformal field theory}\label{sec:dual_cft}

The general arguments of the AdS/CFT correspondence \cite{Maldacena:1997re} imply that string theory on the $AdS_3$ background discussed above should be dual to the 2d ${\cal N}=(4,4)$ superconformal field theory (SCFT) that arises at low energies on a bound state of $Q_1$ D-strings and $Q_5$ D5-branes wrapped on $T^4$. In this section we will discuss this theory when all the R-R scalars vanish. One way to describe this theory is as a sigma model on the moduli space of $Q_1$ instantons of a $U(Q_5)$ theory on $T^4$ (T-duality implies that another way to describe the same low-energy theory is as a sigma model on the moduli space of $Q_5$ instantons of a $U(Q_1)$ gauge theory on a dual $T^4$). The dimension of this moduli space is $4Q_1Q_5+4$, where the first factor comes from the moduli space of $Q_1$ instantons of $SU(Q_5)$ on $T^4$, and the second factor from the Wilson lines of the overall $U(1)$ in $U(Q_5)$ (which do not affect the $SU(Q_5)$ gauge fields, except through global constraints). The corresponding conformal field theory thus has central charge $c=6 Q_1 Q_5 + 6$, consistent at large $Q_1$, $Q_5$ with the central charge following from the supergravity background of the previous section \eqref{eq:newtons}, \eqref{eq:newtons_V}. We will denote $N \equiv Q_1 Q_5$.

In general, this sigma model is a complicated SCFT, and no simple theory that flows to it at low energies is known. In the limit that the volume of $T^4$ becomes infinite, it can be identified (for $Q_5 > 1$) as the conformal field theory describing the Higgs branch of a $U(Q_1)$ ${\cal N}=(4,4)$ supersymmetric QCD theory with one hypermultiplet in the adjoint representation and $Q_5$ hypermultiplets in the fundamental representation (this theory decouples at low energies from the theory of the Coulomb branch \cite{Aharony:1997th,Witten:1997yu}, which corresponds to separating the D-strings from the D5-branes). 

The sigma model on the instanton moduli space is singular for $Q_5 > 1$ when instantons go to zero size; for a single $SU(2)$ instanton on $\mathbb{R}^4$ becoming small, this singularity is locally an $\mathbb{R}^4/\mathbb{Z}_2$ singularity with a vanishing theta angle. At these singularities, the sigma model develops semi-infinite throats (which before taking the low-energy limit connect the Higgs branch to the Coulomb branch \cite{Aharony:1999dw}), with a continuous spectrum of dimensions above a gap of order $Q_5$. 

The moduli space simplifies significantly for $Q_5=1$; in this case all instantons are point-like, so the moduli space of instantons becomes $(T^4)^{Q_1}/S_{Q_1}$. Including the Wilson lines, and noting that in this case $N = Q_1$, the CFT is expected to be a sigma model on
\begin{equation} \label{eq:freeorb}
    (T^4)^{N}/S_{N} \times \hat T^4
\end{equation}
(where the second $\hat T^4$, arising from the $U(1)$ Wilson lines, is inversely related to the first $T^4$). This moduli space has orbifold singularities when instantons come together on the $T^4$ (these are not related to the small-instanton singularities discussed above). A priori the value of the theta angle at these singularities is not clear, but we will review below the arguments that it has the value $\theta=\pi$ corresponding to a free orbifold (rather than the value $\theta=0$ that appeared above).\footnote{Naively this statement contradicts statements made in the previous paragraphs, because we stated that the $(Q_1,Q_5)=(N,1)$ theory is the same as the $(Q_1,Q_5)=(1,N)$ theory (possibly at different values of the moduli), and naively the latter theory of a single instanton in $U(N)$ is expected to have a continuous spectrum because of the small-instanton singularity. However, it turns out that there are actually no smooth $SU(N)$ instantons on $T^4$ \cite{Taubes:1983bk}, so the naive expectation that the local small-instanton singularity looks the same on $T^4$ and on $\mathbb{R}^4$ does not hold in the $Q_1=1$ case. Our arguments imply that the sigma model on the one-instanton moduli space of $U(N)$ on $T^4$ should be given by the free orbifold \eqref{eq:freeorb}, which one would naively obtain from the position of a zero-size-instanton and from the $U(N)$ Wilson lines in this theory, and it would be interesting to confirm this directly.}

Returning to the case of general $Q_5$, as in any ${\cal N}=(4,4)$ SCFT, the super-Virasoro algebra contains $SU(2)_L\times SU(2)_R$ affine Lie algebras of level $Q_1 Q_5 + 1$. In addition, the SCFT described above has $8$ left-moving and $8$ right-moving $U(1)$ affine Lie algebras. $4+4$ of these symmetries have level $Q_1$, and are related to the compact scalars describing the center of mass position of the $Q_1$ instantons (which lives on $T^4$). The other $4+4$ affine Lie algebras have level $Q_5$, and are related to the $U(1) \subset U(Q_5)$ Wilson lines (which live on the dual $\hat T^4$). These symmetries are manifest in the $Q_5=1$ case \eqref{eq:freeorb}. In the dual supergravity description of the previous subsection, if we consider the NS-NS background, the first $8$ $U(1)$'s arise from the NS-NS sector (as the $g_{\mu i}$ and $B_{\mu i}$ components of the metric and $B$ field, where $i$ labels the four coordinates of $T^4$), and the other $8$ $U(1)$'s arise from the R-R sector (by taking suitable components of the R-R 2-form and 4-form potentials). The level of the affine Lie algebras in the CFT is related to the Chern-Simons level of the corresponding $U(1)$ gauge fields on $AdS_3$, and the levels mentioned above agree with type II supergravity on $AdS_3\times S^3\times T^4$.

\subsection{The string worldsheet}\label{sec:string_ws}

There are various methods for studying string theory in the R-R $AdS_3\times S^3\times T^4$ background -- the Green-Schwarz string, the pure spinor string and a hybrid formalism. However, as in other R-R backgrounds, quantization of the string is not yet well-understood in any of these approaches.

The situation is much better in the NS-NS background. This background can be described in the standard RNS formalism (with ${\cal N}=(1,1)$ supergravity on the worldsheet), in which the worldsheet action is a sum of supersymmetric WZW models of $SL(2)$ (describing $AdS_3$ with NS-NS flux) and $SU(2)$ (describing $S^3$ with NS-NS flux), both at level $Q_5$, and a supersymmetric sigma model on $T^4$ (together these give a critical type II superstring). The supersymmetric $SU(2)$ WZW model can be written as a direct sum of a bosonic $SU(2)$ model at level $(Q_5-2)$ and three free fermions, so this description makes sense for $Q_5 \geq 2$, and it was investigated in detail in \cite{Giveon:1998ns,Kutasov:1999xu}.

One unusual property of string theory in these backgrounds is that it has a continuous spectrum. From the point of view of the worldsheet this arises because $SL(2)$ has representations with continuous `spins', while in space-time it is related to ``long strings'' \cite{Seiberg:1999xz} winding around the angular direction of $AdS_3$, that can go all the way to the boundary at a finite cost in energy (so their radial momentum is a continuous parameter). This continuous spectrum can be related to the continuous spectrum arising at small instanton singularities in the SCFT, discussed in the previous subsection.

The NS-NS background may also be described in the hybrid formalism for the worldsheet, where the $AdS_3\times S^3$ part corresponds \cite{Berkovits:1999im} to a super-WZW model on the supergroup $PSU(1,1|2)$ at level $Q_5$ (for a recent review see \cite{Gerigk:2012lqa}). This description makes sense for any integer value of $Q_5$. For $Q_5>1$ it agrees with the RNS formalism discussed above, while for $Q_5=1$ it behaves very differently and, in particular, it does not have a continuous spectrum. It was shown in \cite{Gaberdiel:2018rqv} that the spectrum \cite{Eberhardt:2018ouy,Gaberdiel:2021njm} and correlation functions \cite{Eberhardt:2019ywk,Dei:2020zui,Eberhardt:2020akk,Knighton:2020kuh} of string theory in this background with $Q_5=1$ precisely reproduce those of the symmetric orbifold $(T^4)^{Q_1}/S_{Q_1}$, and we will discuss the precise form of this matching (and its consistency with our discussion in the previous subsection) in more detail below.
Together with the discussion of the previous subsection, and with the fact that this background (like the free orbifold) maps to itself under T-duality, this strongly suggests that string theory on the $(N,1)$ NS-NS $AdS_3$ background with vanishing R-R scalars is dual to the CFT \eqref{eq:freeorb} with $\theta=\pi$, the free orbifold value.

\subsection{Relations between different backgrounds}\label{sec:diffrent_back}

Naively, the backgrounds with different values of $(Q_1,Q_5)$ are distinct, and each of them corresponds to a different SCFT. But in fact, as discussed in detail in \cite{Larsen:1999uk}\footnote{In \cite{Larsen:1999uk} it was assumed that the free orbifold \eqref{eq:freeorb} sits at $Q_5=1$ at a non-zero value of the R-R scalar $\chi$, since at the time it was believed that the $\chi=0$ theory has a continuous spectrum (as it does for higher values of $Q_5$). However, up to moving the subspace associated with the free orbifold to sit at $\chi=0$, the analysis of the dualities relating different backgrounds in \cite{Larsen:1999uk} is still valid, and we review it in this subsection and in section \ref{sec:decoupled_sec}.}, all the backgrounds with the same $N \equiv Q_1 Q_5$ and with mutually prime $(Q_1, Q_5)$ are related by $SO(5,5,\mathbb{Z})$ U-duality transformations of type II string theory on $T^4$. Those $SO(5,5,\mathbb{Z})$ transformations that preserve $(Q_1, Q_5)$ remain as duality symmetries of string theory on $AdS_3\times S^3\times T^4$, and form a subgroup of $SO(4,5,\mathbb{Z})$ (among these, we already mentioned the $SO(4,4,\mathbb{Z})$ T-duality in the NS-NS description). On the other hand, the remaining transformations relate the theories with different values of $(Q_1, Q_5)$; we already described the S-duality transformation that maps NS-NS charges to R-R charges, and in this subsection we focus on transformations between backgrounds with purely NS-NS charges (a more general discussion of these dualities will be given in section \ref{sec:decoupled_sec}). In our descriptions in the previous subsections, the theories with different $Q_1$ and $Q_5$ looked very different, because we focused on the co-dimension 4 subspace of their moduli space where the R-R scalars vanish. However, the duality transformations do not leave this subspace invariant, but rather they relate configurations with vanishing R-R scalars for one value of $(Q_1,Q_5)$ to configurations with non-zero R-R scalars for other values. When the string coupling in one description is small, it is large in all other descriptions, so there is no direct relation between the perturbative string expansions for different values of $(Q_1,Q_5)$. 

\begin{figure}
    \centering
    \includegraphics[width=.88\linewidth]{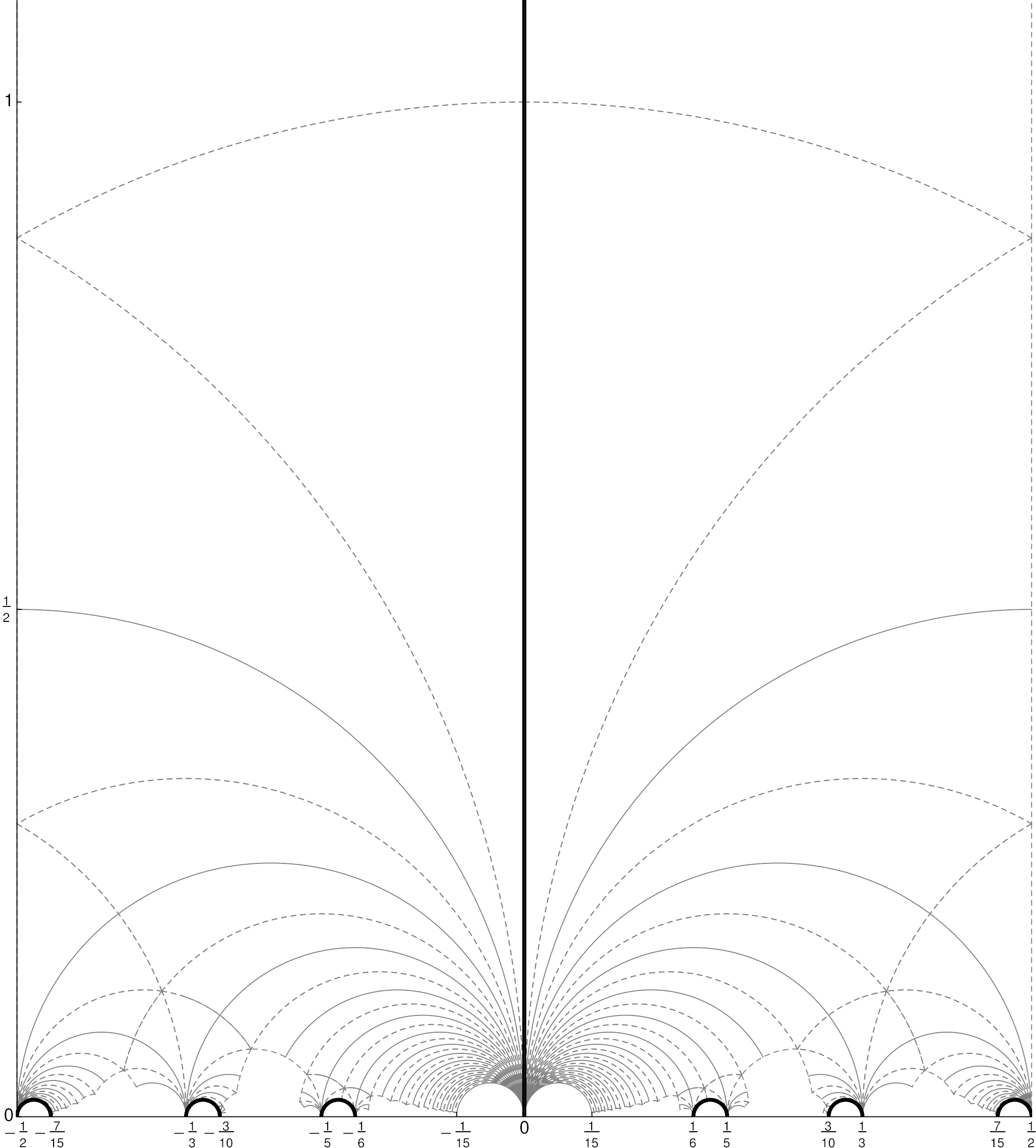}
    \caption{The subspace of the moduli space parameterized by $\tau_\text{RR}=(\chi + i/g_{10})$ in the R-R description of the $Q_1=30$, $Q_5=1$ background, for fixed $T^4$ moduli.
    A fundamental domain under the U-duality subgroup $\Gamma_0(N)$ (that keeps both the charges $(Q_1,Q_5)=(N,1)$ and the $T^4$ normalized metric fixed) is bounded by the grey lines, and divided into different $SL(2,\mathbb{Z})$ fundamental domains. 
    \\\hspace{\textwidth}
    The $(Q_1,Q_5)=(N,1)$ NS-NS background with vanishing R-R scalars sits on the imaginary axis at $\chi=0$, and is weakly coupled close to $\tau= 0$. Its S-dual, the $(N,1)$ R-R background, is weakly-coupled close to $\tau= i \infty$. The two limits are continuously connected to each other as a function of $g_{10}$, drawn in a vertical thick black line, and this line maps to the free orbifold \eqref{eq:freeorb}.
    \\\hspace{\textwidth}
    For every other $(Q_1,Q_5)$ background with $Q_1 Q_5 = N$, there is a line describing its singular ($\chi'=0$) locus, connecting the weakly coupled $(Q_1,Q_5)$ NS-NS and R-R backgrounds. These are the rest of the thick black lines in the figure. Up to the sign of $\chi$, the $(15,2)$ line connects the NS-NS theory at $\tau= 1/2$ to the R-R theory at $\tau=7/15$, the $(10,3)$ line connects the NS-NS theory at $\tau= 1/3$ to the R-R theory at $\tau=3/10$, and the $(6,5)$ line connects the NS-NS theory at $\tau= 1/5$ to the R-R theory at $\tau=1/6$.
    }
    \label{fig:N30_moduli}
\end{figure}

Turning on the R-R scalars corresponds to an exactly marginal deformation of the CFT, so this implies that all the theories with different mutually prime $(Q_1,Q_5)$ but the same $N=Q_1 Q_5$ sit on the same moduli space of exactly marginal deformations. In particular, they can all be described as exactly marginal deformations of the free sigma model on \eqref{eq:freeorb}. This sigma model actually has an 84-dimensional space of exactly marginal deformations. 16 of these are the metric and $B$-field on the $T^4$ in $(T^4)^N/S_N$ (which map to the metric and $B$-field of the $T^4$ in the NS-NS description). 4 additional ones are blow-up modes of the $\mathbb{Z}_2$ singularity of this orbifold. The other 64 exactly marginal deformations take the form $J_i {\bar J}_j$, where $i$ ($j$) go over the 8 left-moving (right-moving) $U(1)$ currents of the SCFT. 16 of these deformations are the metric and $B$ field on the $\hat T^4$ in \eqref{eq:freeorb}, 16 of them can be thought of as changing the metric and $B$ field on the ``center of mass'' $T^4$ in the orbifold, and the remaining 32 mix the $U(1)$ symmetries of the orbifold and those of $\hat T^4$. As discussed in \cite{Larsen:1999uk}, turning on the 4 R-R scalars on $AdS_3$ corresponds in the CFT to a linear combination of turning on the blow-up modes of the orbifold, and specific $J{\bar J}$ deformations mixing the $T^4$ and $\hat T^4$ symmetries. For every mutually prime pair $(Q_1, Q_5)$ there is a co-dimension 4 subspace of the 20-dimensional space of non-$J{\bar J}$ deformations that maps to the $(Q_1,Q_5)$ background with vanishing R-R scalars, and on that subspace the specific $J{\bar J}$ deformations that are related to string theory on $AdS_3$ deform the currents such that there are $4+4$ $U(1)$ currents of level $Q_1$, and $4+4$ additional currents (orthogonal to them) of level $Q_5$. 
In figure \ref{fig:N30_moduli} we show a slice of these subspaces for $N=30$, where we turn on only the string coupling and the 10d R-R scalar field $\chi$ (this determines also the value of the R-R 4-form on the $T^4$); we draw this slice in the language of the R-R background with $Q_1=30$ and $Q_5=1$ (with a fixed $T^4$ shape), and describe the positions of the subspaces mentioned above in this parameterization (a similar figure for $Q_1=6$ appears in \cite{Larsen:1999uk}). This background is mapped to itself under a $\Gamma_0(N)$ subgroup of an $SL(2,\mathbb{Z})\times SL(2,\mathbb{Z})$ subgroup of the U-duality group, which acts on $\tau=\chi+\frac{i}{g_{10}}$ in the same way as the $\Gamma_0(N)$ subgroup of $SL(2,\mathbb{Z})$, and we draw a fundamental domain containing all inequivalent theories (in this slice).\footnote{The number of fundamental domains of $SL(2,\mathbb{Z})$ inside a fundamental domain of $\Gamma_0(N)$ is $N \prod_{p | N, p\ \text{prime}} (1 + p^{-1})$ \cite{schoeneberg2012elliptic}. For $N=30$, which we draw in the figure, it is $72$.} In section \ref{sec:decoupled_sec} we will describe in detail where all the subspaces mentioned above (of $(Q_1,Q_5)$ charges with vanishing R-R fields) sit inside this slice; they are denoted by thick black lines in the figure. Note that the subspaces corresponding to charges $(Q_1,Q_5)$ and $(Q_5,Q_1)$ are identical, since the two descriptions are related (for fixed $T^4$ shape) by a combination of S-duality, T-duality on all $4$ cycles, S-duality and another T-duality on all $4$ cycles.

The other 64 exactly marginal deformations, that are not related to scalar fields in $AdS_3$, are all related to changes in the boundary conditions for the $U(1)^{8}\times U(1)^8$ CS gauge fields on $AdS_3$ (these can be written as 8 copies of a $\frac{k}{4\pi} \int A \wedge dB$ theory, where $k$ is equal to either $Q_5$ or $Q_1$, depending on the gauge field). For a given choice of complex coordinates in the CFT, they can be thought of as choosing which $8$ bulk gauge fields have a boundary condition in which their $z$-component is fixed at the boundary (so that they correspond to anti-holomorphic currents), while the other $8$ gauge fields have their ${\bar z}$ component fixed at the boundary.

Note that for $Q_5>1$ with vanishing R-R scalars the CFT contains a long-string sector with a continuous spectrum, which is far from evident in our description of the CFT as a deformation of the symmetric orbifold. The subsector of the CFT describing the long strings is believed to itself be described as an $S_{Q_1}$ symmetric orbifold \cite{Seiberg:1999xz,Argurio:2000tb,Eberhardt:2019qcl,Dei:2019osr}, and it would be nice to understand how this is related to the picture described above.\footnote{It was suggested in \cite{Eberhardt:2021vsx} that perhaps even the full CFT has a description as a deformed symmetric orbifold of this type.}

There is another subspace of the moduli space of the deformed free orbifold which is singular, which is the subspace where the metric is given by \eqref{eq:freeorb} but the theta angle on the vanishing 2-cycle at the $\mathbb{Z}_2$ orbifold singularity of \eqref{eq:freeorb} is taken to $\theta=0$ (instead of its value $\theta=\pi$ at the free orbifold point \cite{Aspinwall:1995zi}). This subspace is not the same as the singular subspaces that are weakly coupled NS-NS-background strings; we believe that in the R-R background of figure \ref{fig:N30_moduli} it sits at $\chi = \pm 1/2$, on the boundary of the fundamental domain.

\subsection{Questions}

The picture described in this section raises 3 questions:
\begin{enumerate}
    \item String theory in NS-NS $AdS_3$ backgrounds is believed \cite{Kutasov:1999xu,Giveon:2001up,Kim:2015gak,Eberhardt:2020bgq,Eberhardt:2021jvj} to describe not a background with fixed $Q_1$, but rather a grand-canonical ensemble of theories with different $Q_1$ (and the same $Q_5$).
    How is this consistent with the picture described above, and how is it consistent with the detailed matching of the correlation functions of the $Q_5=1$ string to the free orbifold \eqref{eq:freeorb}?
    \item For $Q_5=1$ we mentioned that the NS-NS perturbative expansion matches the $1/N$ expansion of the free orbifold \eqref{eq:freeorb}. However, the orbifold is free for any value of the moduli of the torus, while \eqref{eq:ns_ns_desc} implies that the $(N,1)$ NS-NS string theory becomes strongly coupled once $v_4 > N$. Moreover, for $v_4 \gg N$ it seems that the same theory should have a different weakly coupled description, as string theory on the R-R background. How is this consistent?
    \item The SCFT \eqref{eq:freeorb} (dual to string theory with $Q_5=1$) has a decoupled $\hat T^4$ sector, which did not show up in the mapping of this string theory to the free orbifold. Moreover, this sector is still essentially decoupled from the rest of the SCFT even after we deform it by $J{\bar J}$ deformations to go to other values of $(Q_1, Q_5)$ (one can think of these deformations as just modifying the energies of states with given charges). How is this decoupled sector realized in string theory?
\end{enumerate}

The answers to these questions will be discussed in the next three sections of the paper, respectively. The three sections are almost completely independent of each other, so readers who are just interested in one question can jump directly to the relevant section.

\section{String theory and the grand-canonical ensemble}\label{sec:st_grand_can}

The near-horizon limit suggests that (the non-perturbative completion of) string theory on $AdS_3\times S^3\times T^4$ with fixed fluxes $Q_1$ and $Q_5$ should be equivalent to the conformal field theories discussed above, at any point in their moduli space. In principle, one can consider instead of the quantum gravity theory (and its dual CFT) with some fixed flux $Q$, the grand-canonical ensemble defined as the sum over the theories with flux $Q$, weighted by $e^{-\mu Q}$. In the bulk this can be thought of as choosing a different boundary condition for the 2-form field potential on $AdS_3$, for which the integral over $AdS_3$ of its field strength is equal to $Q$. 

Somewhat surprisingly, it turns out that perturbative string theory on the NS-NS background with NS5-brane flux $Q_5$ computes such a grand-canonical ensemble with respect to $Q_1$, rather than being dual to a specific conformal field theory.\footnote{Note that perturbative string theory in the R-R background is believed to correspond to fixed fluxes; in particular string theory in this background has a T-duality symmetry exchanging $Q_1$ and $Q_5$.} This was first understood in the RNS formalism for $Q_5 \geq 2$ \cite{Kutasov:1999xu,Giveon:2001up,Kim:2015gak}, and it was then claimed to be the case also in the hybrid formalism for $Q_5=1$ \cite{Eberhardt:2020bgq,Eberhardt:2021jvj}. In this section, we explain in detail how the grand-canonical ensemble for $Q_5=1$ is consistent with the relation between free orbifold correlation functions and perturbative string theory.

\subsection{A brief review of string theory in the NS-NS background}

In this subsection we review the relevant information about string theory on $AdS_3\times S^3\times T^4$ with a purely NS-NS background. 

As reviewed in section \ref{sec:sugra}, when we take the near-horizon limit of $Q_1$ fundamental strings and $Q_5$ NS5-branes wrapped on $T^4$, we obtain this $AdS_3$ background, with a fixed value for the six-dimensional string coupling (related to the vacuum expectation value of the dilaton field) given by $g_6'^2 = Q_5 / Q_1$. So from this point of view, we would expect string theory on the NS-NS background to have a fixed value of $Q_1$ and not to have any continuous NS-NS parameters beyond the moduli of the $T^4$.

However, if we directly study string theory on the NS-NS $AdS_3$ background, either in the RNS formalism (for $Q_5 \geq 2$) \cite{Giveon:1998ns,Kutasov:1999xu,Giveon:2001up,Kim:2015gak} or in the hybrid formalism \cite{Eberhardt:2020bgq,Eberhardt:2021jvj}, we find a rather different picture.
The worldsheet action depends explicitly on $Q_5$, but the parameter $Q_1$ does not appear in it. On the worldsheet one can have as usual a continuous parameter $g_s$ related to the coefficient of $\int d^2\sigma \sqrt{{\rm det}(h)} R[h]$ where $h$ is the worldsheet metric, which weighs connected string diagrams arising from genus $g$ surfaces by $g_s^{2g-2}$. Unlike in most other string theories, here this parameter is not related to the expectation value of any field in space-time, but it is still present on the worldsheet.

In addition, when computing from the worldsheet the central charge appearing in the space-time operator product expansion (OPE) of two CFT energy-momentum tensors or NS-NS $U(1)$ currents, one finds that it is given by $c = 6 Q_5 I$ and by $k = I$, respectively, where $I$ is a specific operator on the worldsheet (naively it depends on a position on the boundary of $AdS$ space, but in fact this dependence is trivial) \cite{Giveon:1998ns}. If the string theory corresponded to the CFT suggested by the near-horizon limit, we would expect to have $Q_1$ appearing in these equations instead of $I$, but instead the operator $I$ appears, whose correlation functions are not equivalent to replacing it by a constant. The correlation functions of this operator are not arbitrary, but are fixed by the worldsheet theory. The fact that $I$ is not a constant prevents the interpretation of string theory as dual to a specific CFT with a given central charge, and from having the expected space-time factorization properties.

The interpretation of this property was suggested in \cite{Kim:2015gak,Eberhardt:2020bgq, Eberhardt:2021jvj}. Since $I$ is a vertex operator on the worldsheet, it is natural to turn on a coupling $\mu$ for it, like for any other vertex operator. Then, naively one obtains a string theory labelled by two continuous parameters, $g_s$ and $\mu$. However, the properties of correlation functions of $I$ imply a specific dependence of correlation functions on $\mu$ \cite{Giveon:2001up,Kim:2015gak}, and this dependence is such that its effect can be swallowed into a rescaling of the string coupling $g_s$ and a $\mu$-dependent normalization of the worldsheet vertex operators. So, in this normalization of the vertex operators, string theory on $AdS_3$ depends just on a single continuous parameter $g_s$. It was then suggested that this string theory should be identified with a grand canonical ensemble of the $(Q_1,Q_5)$ CFTs reviewed in section \ref{sec:review}, with the same $Q_5$ but with different values of $Q_1$, weighted by $p^{Q_1}$, with some specific relation between the continuous parameters $p$ and $g_s$. It was argued in \cite{Kim:2015gak} that one can obtain the CFT with fixed $Q_1$ by a Legendre transform of the string theory. In this section, following the analysis of $Q_5=1$ partition functions in \cite{Eberhardt:2021jvj}, we describe the precise relation between the grand canonical ensemble of CFTs and the dual string theory. For $Q_5=1$ we argue that the precise way to obtain fixed-$Q_1$ CFT correlation functions from string theory is more complicated than a Legendre transform (though it agrees with it at leading order), and we conjecture that this should be true also for $Q_5>1$.

\subsection{The genus expansion for symmetric orbifolds} \label{sec:genus_exp}

Before introducing the grand-canonical ensemble, it is illuminating to understand why the fixed $N$ orbifold theory \eqref{eq:freeorb} does not behave as a string theory. Specifically, it does not have a proper genus expansion. Even though we focus on the $(T^4)^N/S_N$ orbifold in this paper, our analysis of the free orbifold in this section is relevant for any symmetric orbifold.

We begin by briefly reviewing the construction of the genus expansion for the (fixed $N$) $(T^4)^N/S_N$ symmetric orbifold, following   \cite{Lunin:2000yv,Pakman:2009zz} (see also \cite{Lunin:2001pw,Pakman:2009ab,Giribet:2007wp,Eberhardt:2019ywk}). The Hilbert space of the symmetric orbifold at low energies and large $N$ can be understood as a Fock space of single-cycle-twist operators, divided into different $w$-sectors. The basic operator for a given such sector $1 < w \leq N$ is the  $w$-cycle twist operator, defined as
\begin{equation} \label{eq:my_norm}
    \sigma_w(z) = \frac{1}{w (N-w)!} \sum_{h\in S_N} \sigma_{h^{-1} (1,...,w) h}(z),
\end{equation}
where $(1,...,w)$ is the cyclic permutation of the first $w$ elements, and $\sigma_g(z)$ is the twist operator with holonomy $g\in S_N$ permuting the $N$ copies as one goes around it.
The sum over $S_N$ ensures the gauge invariance of the operator. 
Our choice of normalization may appear non-standard, but as we will see below, this is the correct normalization for comparison with string theory vertex-operators. It is mathematically natural, as the denominator is the stabilizer size of $(1,...,w)$ in $S_N$ (by $S_N$ conjugations), so that each different permutation is counted in \eqref{eq:my_norm} with weight $1$.
In this section we will consider only correlation functions of $\sigma_w$'s, but the extension to more general single-cycle operators (with additional CFT excitations in addition to the twist operator) and to untwisted-sector operators (which have $w=1$) is straightforward.

The insertion of $\sigma_w(0)$ can be understood as cutting a hole around $z=0$, with a cyclic gluing of $w$ copies of $T^4$ around the boundary circle (different ones for each term in \eqref{eq:my_norm}), and a trivial gluing for the rest of the $(N-w)$ copies.  A correlation function of $\sigma_w(z)$'s will get contributions from consistent gluings between the insertions. Each such consistent gluing between the $N$ copies gives a non-trivial covering of the original CFT manifold. The contribution of each gluing to the correlation function is proportional to the partition function of the $T^4$ CFT on the covering space. For this reason, the covering space is believed \cite{Pakman:2009zz} (and later also shown \cite{Eberhardt:2019ywk,Dei:2020zui,Eberhardt:2020akk,Knighton:2020kuh,Knighton:2024ybs}) to be identified with the string worldsheet.

We begin with the two-point function of two $\sigma_w$'s on the plane (which can be conformally mapped to the sphere), given by \cite{Pakman:2009zz}
\begin{equation}\label{eq:sigma_2p}
    \langle \sigma_{w}(z) \sigma_{w'}(0)\rangle = \frac{N!}{w (N-w)!} \frac{\delta_{w,w'}}{|z|^{2\Delta_{w}}},
\end{equation}
with $\Delta_{w}=\frac{c}{24}(w-1/w)$ (where in our case $c=6$ is the central charge of the $T^4$ SCFT).
In this case all consistent gluings are related by a gauge transformation, and the topology of the gluing is of a sphere with two insertions of degree-$w$ branch points (more precisely, this is the topology of the non-trivial part of the correlation function which involves $w$ copies of the $T^4$, ignoring all other copies). We thus understand \eqref{eq:sigma_2p} as a single worldsheet diagram contribution with genus zero.
The combinatorial factor of the diagram can be understood as follows. In the normalization we choose, each gluing is summed over once, and we only need to count how many different permutations are gauge-equivalent to $(1,...,w)$. This is given by $N!$ divided by the size of the stabilizer of $(1,...,w)$, which is $w (N-w)!$.

In a general diagram contributing to a correlation function on the plane, there will be $N_c$ copies of $T^4$ which are involved in non-trivial permutations, while the other $(N-N_c)$ copies may be viewed as ``vacuum diagrams''. We will call a diagram ``connected'' if all the $N_c$ copies are permuted non-trivially (rather than having some of them only permuted among themselves).
By the Riemann–Hurwitz formula, in connected diagrams $N_c$ is related to the genus $g$ of the covering space by
\begin{equation}\label{eq:N_c_def}
    N_c = 1-g + \frac{1}{2}\sum_{j=1}^n (w_j -1).
\end{equation}
The contribution from such connected diagrams goes as
\begin{equation}\label{eq:conn}
    \langle \sigma_{w_1}(z_1) ... \sigma_{w_s}(z_s)\rangle_\text{conn} = 
    \sum_{g=0}^{g_\text{max}} \frac{N!}{N_c (N-N_c)!} \cdot (...),
\end{equation}
where $(...)$ stands for the contribution of diagrams with the appropriate value of $N_c$, given in terms of the path-integral over the appropriate branched covering, divided by the appropriate vacuum path-integral \cite{Lunin:2000yv}. 
For a given $N_c$, the combinatorial factor in \eqref{eq:conn} is simply the orbit size of the cyclic permutation $(1,...,N_c)$ in $S_N$; in a specific diagram, after choosing the $N_c$ participating sheets and their cyclic ordering, there is just one permutation from \eqref{eq:my_norm} which contributes for each operator. Here we are interested in the large $N$ limit in which the $w_i$ are kept fixed; for finite values of $N$ there is an extra restriction that $N_c \leq N$.

In the literature \cite{Lunin:2000yv,Pakman:2009zz} one usually defines a normalized version $\tilde \sigma_w$ of the twisted operators \eqref{eq:my_norm}, in which their two-point function \eqref{eq:sigma_2p} is normalized to one. Then one can ask about the leading large $N$ behavior of a given connected diagram to \eqref{eq:conn}. The result from genus $g$ diagrams is
\begin{equation} \label{eq:N_count}
    \langle \tilde \sigma_{w_1}\cdots \tilde \sigma_{w_n}\rangle_{\text{conn},g} \propto \frac{N!}{N_c (N-N_c)!}\prod_{j=1}^n \sqrt{\frac{w_j (N-w_j)!}{N!}} \sim N^{N_c - \frac{1}{2}\sum_{j=1}^n w_j} = N^{1-g-n/2}.
\end{equation}
This result looks very appealing, after comparing it to the general structure expected in string theory,
\begin{equation}\label{eq:string_exp}
    \langle O_1 \cdots O_n \rangle_\text{conn} = \sum_{g=0}^{\infty} g_s^{2g-2+n}\cdot \sum_{\text{Diagrams with genus } g} (....).
\end{equation}
A naive comparison gives $g_s^2 = 1/N$, with an identification between string theory diagrams and the symmetric orbifold's sum over branched coverings.

However, this identification actually works only at leading order in $1/N$. We will give two different reasons to suspect the identification. The first reason, which was noticed already in \cite{Pakman:2009ab}, is that the right-hand side of \eqref{eq:N_count} describes only the leading $1/N$ behavior from a given genus, and receives $1/N$ corrections from expanding the product in the middle. Thus, at subleading order in $1/N$, a given correlator will receive contributions from higher genus diagrams, but also from the subleading corrections to the planar diagrams, and we find
\begin{equation}
    \langle O_1 \cdots O_n \rangle_\text{conn} = g_s^{n-2}\cdot (\text{genus 0}) +  g_s^{n}\cdot (\text{genus 1}+\# \cdot \text{genus 0}) + \cdots \ .
\end{equation}
This is not what we expect in \eqref{eq:string_exp}. Note that because of the structure of \eqref{eq:N_count}, it is not possible to fix this by redefining $g_s$ and/or the operator normalizations at higher orders in $1/N$.

The second problem with identifying $g_s^2=1/N$ arises for disconnected diagrams. For simplicity, consider the four-point function of two $w_1$ and two $w_2$ twist operators. Diagrammatically the four-point function has a leading disconnected piece, with $w_1$ sheets permuted among themselves and $w_2$ other sheets permuted among themselves. Following the string theory picture, one would expect this contribution to be equal to the product of the two two-point functions. However, for a given $N$, the combinatorial factor associated with this disconnected piece also accounts for the fact that the $w_1$ and $w_2$ cycles should not overlap. The exact ratio between the disconnected piece and the product of the connected diagrams is thus
\begin{equation}\label{eq:ratio}
    \frac{\langle \sigma_{w_1}(z_1)\sigma_{w_1}(z_2)\sigma_{w_2}(z_3)\sigma_{w_2}(z_4)\rangle_\text{disc}}{\langle \sigma_{w_1}(z_1)\sigma_{w_1}(z_2)\rangle \langle \sigma_{w_2}(z_3)\sigma_{w_2}(z_4)\rangle} = \frac{(N-w_1)!(N-w_2)!}{N! (N-w_1-w_2)!} = 1 +O(1/N),
\end{equation}
and is not exactly equal to $1$. As a result, at fixed $N$, correlation functions do not obey the expected ``cluster decomposition" of string theory beyond the leading order in $1/N$.\footnote{We stress that the CFT still maintains standard field-theory cluster decomposition at fixed $N$. It is only the string theory interpretation that breaks. It was noticed in \cite{Kim:2015gak} that the string theory dual for $Q_5>1$ does not obey field-theory cluster decomposition, due to the Legendre transform; we will comment further on $Q_5>1$ below.}

\subsection{The grand canonical ensemble} \label{sec:grand}

To solve this issue we would like to work not with fixed $N$ but rather in the grand canonical ensemble with a chemical potential for $N$. As we reviewed at the beginning of this section, this is generally expected for string theory on $AdS_3$ with NS-NS backgrounds. It was suggested in \cite{Kim:2015gak} that the string theory at $Q_5>1$ is related to a fixed CFT by taking a Legendre transform with respect to the coefficient of the operator $I$. A slightly different suggestion was made in \cite{Eberhardt:2021jvj} for $Q_5=1$, that string theory and the dual CFT are related by a Laplace transform, so that the string theory corresponds to a grand canonical ensemble of the $(Q_1,Q_5)$ theories with the same NS-NS moduli, with a chemical potential for $Q_1$. In this subsection we analyze this suggestion in detail for $Q_5=1$, generalizing the analysis of the partition function in \cite{Eberhardt:2021jvj} to general correlation functions. In this case we just have a grand canonical ensemble of \eqref{eq:freeorb} with respect to $N$; since this does not affect the ${\hat T}^4$ factor, we will ignore it here and return to it in section \ref{sec:decoupled_sec}.

Denoting by $Z_N[J]$ the partition function of the $\text{Sym}_N(T^4)$ CFT on some manifold with sources $J$ for various operators, the grand canonical partition function is defined as\footnote{As we will discuss below, $p$ is related to the string coupling. In general it is not clear if the sum \eqref{eq:grand} converges, and correspondingly we do not expect string perturbation theory to converge but just to be an asymptotic series. Our statement is that the asymptotic series on the two sides of the duality are the same. In the special case of the free orbifold CFT on a sphere we will see that the sum does converge (at least for any finite number of operator insertions).}
\begin{equation} \label{eq:grand}
    Z_p[J] = \sum_{N=0}^\infty p^N Z_N[J],
\end{equation}
with $Z_0[J]=1$. Sources for untwisted sector operators exist for any $N$, while sources for twisted operators start appearing at some $N_{min}$ that depends on the twisted sector. Note that we are summing here $Z_N[J]$ rather than $Z_N[J]/Z_N[0]$ (which would be the generating function for correlation functions in the $\text{Sym}_N(T^4)$ CFT), since already the vacuum partition function (discussed in \cite{Eberhardt:2021jvj}) is non-trivial when the CFT lives on higher-genus manifolds.

Simplifying to the case where we turn on sources just for the leading twist operators in each twisted sector of the orbifold, the $\text{Sym}_N(T^4)$ generating function is given schematically by\footnote{We include here the $w=1$ case even though $\sigma_1(z)$ is trivial, because the same formulas will be relevant also for sources for any other operators from the same sectors.}
\begin{equation} \label{eq:gen_func}
    Z_N[J] = \frac{1}{N!} \int D^N X e^{-S[X]} \exp\left(\sum_{w=1}^\infty \int d^2 z J_w(z) \sigma_w(z)\right).
\end{equation}
The expansion of $Z_N[J]$ generally includes disconnected coverings. Some of the connected components include operator insertions and others, the vacuum diagrams, do not. The normalization \eqref{eq:my_norm} exactly allows us to compute the combinatorial factor from each component on equal footing. Let us denote by $N_1,...,N_k$ and $n_1,...,n_k$ the number of copies of $T^4$ of the different connected components (which can differ in the number of copies and/or in the operator insertions) and their multiplicity, respectively, so that $\sum_{i=1}^k N_i n_i = N$. The generalization of \eqref{eq:conn} is given by
\begin{equation}\label{eq:gen_Z}
    Z_N[J] = \sum_{n=0}^\infty \frac{(\sum_{w=1}^\infty \int d^2 z J_{w}(z))^n}{n!}
    \sum_{\substack{\text{Disconnected} \\ \text{diagrams}\\ \sum_{i=1}^k n_i N_i = N}}
    \prod_{i=1}^k \frac{1}{n_i ! N_i^{n_i}} (Z_\text{covering}^{(i)})^{n_i},
\end{equation}
with $Z_\text{covering}^{(i)}$ the partition function coming from the $i$-th connected component (such that the product of the partition functions contains also insertions of $n$ $\sigma_{w_j}(z_j)$ operators)\footnote{We are not writing down explicitly the dependence of the CFT partition function on the metric (some such dependence is required by the conformal anomaly), and, if it is on a manifold of genus $G \geq 1$, on the spin structure. The covering spaces inherit the metric and spin structure in an obvious way from those of the original manifold.}. We stress again that some of these components include no operator insertions and correspond to vacuum diagrams. For example, in \eqref{eq:conn} there's a single $n_1=1, N_1=N_c$ connected diagram with all the insertions, and $n_2=N-N_c$ vacuum diagrams with $N_2=1$ each. For given operator insertions, not all the possible $N_i$'s and $n_i$'s are included in the sum, as only some will correspond to a possible diagram. For example, only $N_i=1$ diagrams are possible when all $w_j=1$. Of course, for some values of $\{w_j\}$ there are no covering spaces at all if the corresponding permutations cannot be multiplied to give the identity permutation (and in particular there are no diagrams if any $w_j > N$).

Importantly, the only dependence on $N$ in \eqref{eq:gen_Z} is in the definition of the sum over $N_i, n_i$. Considering now the grand-canonical ensemble \eqref{eq:grand}, the sum over the diagrams is no longer constrained by $N$. Notice that as \eqref{eq:my_norm} are cyclic, each insertion will appear in a single connected component.\footnote{The fact that only for single-cycle operators the grand-canonical partition function exponentiates further supports their identification as single-string states.}
As a result, the now unconstrained sums over $n$ and $n_i$ nicely exponentiate into connected components
\begin{equation}\label{eq:grand_Z}
    Z_p[J] = \exp\left\{
    \sum_{n=0}^\infty \frac{(\sum_{w=1}^\infty \int d^2 z J_{w}(z))^n}{n!}
    \sum_{\substack{\text{Connected} \\ \text{diagrams}}} \frac{p^{N_c}}{N_c} {Z}_\text{covering}^{\{w_j\}} \right\},
\end{equation}
with $N_c$ the number of $T^4$'s in the connected covering, and $Z_\text{covering}^{\{w_j\}}$ the corresponding partition function in the presence of the $n$ $\sigma_{w_j}(z_j)$ operators.
In this expression, the vacuum diagrams appear in the $n=0$ term in the exponent. The vacuum partition function is known to exponentiate nicely in the grand-canonical ensemble \cite{Dijkgraaf:1996xw,Bantay:2000eq}. Using the normalization \eqref{eq:my_norm}, \eqref{eq:grand_Z} is a generalization of that statement which incorporates operator insertions.\footnote{It would be interesting to study extensions of \eqref{eq:grand_Z} to incorporate defects \cite{Knighton:2024noc}, and other permutation groups \cite{Haehl:2014yla,Belin:2015hwa}.}

Up to now our discussion was valid for the CFT on any manifold, but let us now consider the case of the CFT on the sphere (higher genus surfaces will be discussed in the next subsection). In this case there is a single vacuum diagram in $Z_N$, which gives $Z_N[0] = Z_1[0]^N / N!$. Thus, the grand-canonical partition function for the sphere can also be written as
\begin{equation}\label{eq:grand_sphere_Z}
    Z_p^\text{sphere}[J] = \exp\left\{
    \sum_{n=1}^\infty \frac{(\sum_{w_n=1}^\infty \int d^2 z J_{w}(z))^n}{n!}
    \sum_{\substack{\text{Connected} \\ \text{non-vacuum}}} \frac{p^{N_c}}{N_c} {Z}_\text{covering}^{\{w_j\}} + p \cdot Z_1[0]
    \right\}.
\end{equation}

Unlike the fixed $N$ ensemble discussed above, the exponential form of \eqref{eq:grand_Z} has a nice interpretation as a perturbative string theory. What is the corresponding string coupling?
Redefining our operators by
\begin{equation} \label{eq:op_redef}
    \sigma_w' = p^{-w/2} \sigma_w,
\end{equation}
and using \eqref{eq:N_c_def}, the grand-canonical generating function for $\sigma_w'$ takes the form
\begin{equation}\label{eq:grand_Z_prime}
    Z_p[J'] = \exp\left\{
    \sum_{n=0}^\infty \frac{(\sum_{w=1}^\infty \int d^2 z J_{w}'(z))^n}{n!}
    \sum_{\substack{\text{Connected} \\ \text{diagrams}}} \frac{p^{1-g-n/2}}{N_c} {Z}_\text{covering}^{\{w_j\}} \right\}.
\end{equation}
Comparing with \eqref{eq:string_exp} gives exactly the string theory genus expansion with
\begin{equation} \label{eq:string_coupling_p}
    g_s^{-2} \propto p.
\end{equation}
The problems discussed in the previous subsection disappear when going to the grand-canonical ensemble.

Note that a priori one may expect string theory to compute a grand-canonical generating function for the correlation functions, where the coefficient of $p^N$ in the expansion of an $n$-point correlation function would be the correlation function in the $N$'th CFT. Instead, we find that string theory computes correlation functions that are derivatives of $Z_p[J]$ divided by $Z_p[0]$, and these are not directly related to correlators for specific values of $N$ (which are derivatives of $Z_N[J]$ divided by $Z_N[0]$). To obtain correlators with a specific $N$ one has to separately inverse-Laplace-transform the correlation functions and the vacuum diagrams of the string theory.
In particular, it seems that one should view the grand canonical ensemble not as a sum over different independent theories, but more like summing over some extra `particle number' charge in a given theory, that is present also in the vacuum diagrams.\footnote{This suggests in particular that the string theory dual to the CFT on a disconnected surface should also correspond to a product of the corresponding $Z_p$'s, rather than the naive product of the corresponding $Z_N$'s, as discussed in \cite{Eberhardt:2021jvj}.}

To make the matching to string theory more precise, we can define an operator ``$I_\text{CFT}$'' that would couple to $\log(p)$, and that should be related to the operator $I$ discussed above on the worldsheet \cite{Giveon:1998ns,Giveon:2001up,Eberhardt:2021jvj}.
For the free orbifold ($Q_5=1$) this can be identified with ``$N$" inside the sum over $N$ in \eqref{eq:grand}. The VEV of $I_\text{CFT}$ for the CFT on the sphere (from vacuum diagrams), related to string theory on Euclidean $AdS_3$, is given by 
\begin{equation}
    \langle I_\text{CFT} \rangle = \partial_{\log (p)}\log (Z_p[0]) = p Z_1[0].
\end{equation}
This agrees with the expectation that for large $p$ (corresponding to weak string coupling) the sum over $N$ should have a saddle point with $N \propto p$, with a smaller and smaller variation as $p$ increases. Note that because of the conformal anomaly, $Z_1[0]$ depends on the radius of the sphere that the CFT lives on as $Z_1[0] \propto R^2$. So the relation between $p$ and the typical value of $N$ depends on this radius (see the related discussion in \cite{Eberhardt:2023lwd}). However, the relation between the string coupling and this typical value does not depend on the radius. This can be seen by rewriting $p^{N_c} {Z}_\text{covering}^{\{w_j\}}$ in \eqref{eq:grand_sphere_Z} as $(p Z_1[0])^{N_c} ({Z}_\text{covering}^{\{w_j\}} / Z_1[0]^{N_c})$, where the factor in the second parenthesis is independent of the radius, suggesting that a more precise version of \eqref{eq:string_coupling_p} is $g_s^{-2} \propto p Z_1[0]$.

To find the expectation value of $I_\text{CFT}$ for a given connected diagram, we can compute the derivative of the generating function $\log (Z_p[J])$ by $\log (p)$. Directly from the form of $\log (Z_p[J])$ in \eqref{eq:grand_Z} we get
\begin{equation}\label{eq:I_conn}
    \left\langle I_\text{CFT} \ \prod_{j=1}^n \sigma_w(z_j) \right\rangle_{\text{conn},g} = N_c \cdot \left\langle \prod_{j=1}^n \sigma_w(z_j) \right\rangle_{\text{conn},g}.
\end{equation}
This agrees with the arguments of \cite{Eberhardt:2021jvj} for correlation functions of $I$ in the corresponding string theory.
 
 In \cite{Kim:2015gak} it was noted that the string theory disconnected 4-point function (for $Q_5 > 1$) does not satisfy the naively expected field-theoretic cluster decomposition, due to the appearance of the central element $I$ in the appropriate channel. Similarly, for $Q_5=1$ where we can compute the correlation functions from $Z_p$, what would naively be identified as a connected correlation function in the dual CFT is given by
 \begin{equation}
 \begin{split}
    \langle \sigma_{w_1}(z_1)\sigma_{w_1}&(z_2)\sigma_{w_2}(z_3)\sigma_{w_2}(z_4)\rangle_p
     - \left\langle \sigma_{w_1}(z_1)\sigma_{w_1}(z_2)\right\rangle_p \left\langle\sigma_{w_2}(z_3)\sigma_{w_2}(z_4)\right\rangle_p\\
     & = \frac{\partial_{J_{w_1}}^2\partial_{J_{w_1}}^2 Z_p[0]}{Z_p[0]} - \frac{\partial_{J_{w_1}}^2 Z_p[0]}{Z_p[0]}\frac{\partial_{J_{w_1}}^2 Z_p[0]}{Z_p[0]}\\
     & = \left(1-\exp(-p \cdot Z_1[0])\right) \cdot \left\langle \sigma_{w_1}(z_1)\sigma_{w_1}(z_2)\right\rangle \left\langle\sigma_{w_2}(z_3)\sigma_{w_2}(z_4)\right\rangle + \left(\text{Connected}\right).
\end{split}
\end{equation}
The first term on the right-hand side does not vanish when the distance between the pairs of operators becomes large, as expected for a standard CFT. Expanding in orders of $p$, the sum can be interpreted as the OPE contribution of $(I-\langle I \rangle)^k$ operators.

\subsection{Higher genus partition functions} \label{sec:higher_genus_CFT}

Equation \eqref{eq:grand_Z} holds also for the CFT on higher genus Riemann surfaces (which is expected to be dual to string theory on orbifolds of $AdS_3$). The main difference is that already without any operator insertions, we now have a sum over holonomies in $S_N$ around the various cycles of the Riemann surface (subject to a global consistency condition), and a similar sum over holonomies appears in the correlation functions of vertex operators. So, in \eqref{eq:grand_Z} the sum over coverings should also allow different $S_N$ holonomies (already for the $n=0$ term). The definition of the $N_c$ `participating' sheets now involves both the sheets that are permuted by vertex operators, and the ones that are permuted by non-trivial holonomies (it is still the degree of the covering). The splitting into `connected' contributions (which are linked by some permutation, either from a vertex operator or a holonomy) is the same. Because of the sum over the holonomies, there is no longer a simple relation between $Z_N[0]$ and $Z_1[0]$.

Surprisingly, $Z_p$ has peculiar properties for higher genus. Let us denote the genus of the CFT manifold by $G$, and assume $G>1$. To repeat the calculation of the string coupling, we can use the generalization of \eqref{eq:N_c_def} to general genus $G$
\begin{equation}\label{eq:N_c_G}
    (1-G) N_c = (1-g) + \frac{1}{2} \sum_{j=1}^n (w_j-1).
\end{equation}
Unlike the $G=0$ case where we had contributions from $g=0,\cdots,g_{max}$, for $G>1$ the worldsheet genus $g$ can range from $g_{min} = G + \sum_j (w_j-1)/2$ to values of order $N$.

Repeating the same redefinition \eqref{eq:op_redef} of the operators,
the generating function for $\sigma_w'$ now gives the string coupling identification
\begin{equation}
    p \propto g_s^{2G-2},
\end{equation}
as also found by \cite{Eberhardt:2021jvj}. Note that while for $G=0$ small coupling corresponded to large $p$ (and therefore large $N$), for $G>1$ the string coupling appears to become large for large $p$. To understand the reason behind this, we repeat the calculation of $\langle I_\text{CFT} \rangle$ for general $G$,
\begin{equation}
    \langle I_\text{CFT} \rangle_G 
    = \sum_{N_c=1} p^{N_c} Z_{\text{Vac},N_c}^{(G)}
    = \sum_{g} g_s^{2g-2} Z_{\text{Vac},g}^{(G)},
\end{equation}
with $Z_{\text{Vac}}^{(G)}$ the corresponding vacuum diagram.
For the sphere $G=0$, we only had the sphere $g=0$ diagram, which gave the expected leading behavior $\langle I_\text{CFT} \rangle \sim g_s^{-2}$. For $G>1$ one would expect a similar relation to arise from the sphere diagram. However
 because in this theory the worldsheet is localized to be a covering of the CFT manifold, the minimal worldsheet genus is $g=G$, so we find (for small $g_s$) the peculiar relation
\begin{equation}
    \langle I_\text{CFT} \rangle_G \sim p \sim g_s^{2G-2}.
\end{equation}
This relation explains why the large $p$ (or $N$) limit does not have the same relation to the string coupling for different $G$'s. We emphasize that in any correlation function, \eqref{eq:N_c_G} forbids tree-level $g=0$ diagrams for $G\ge 1$. This is a unique property of the $Q_5=1$ string theory, due to the localization of the worldsheet to the boundary, and we do not expect it (and the related properties discussed in this subsection) to persist for $Q_5>1$.

An even more peculiar case is the CFT on the torus $G=1$ (related to the string theory on thermal AdS, which is an orbifold of $AdS_3$). In this case \eqref{eq:N_c_G} degenerates, so that the worldsheet genus is completely determined by the operator insertions (with $g=1$ for vacuum diagrams), independently of $N_c$. For that reason we can set $p=1$, and `mimic' a genus expansion by simply appropriately defining a normalized $\sigma'_w$.

\subsection{Final comments}

It is natural to guess that a similar relation between string theory and the CFT applies also to $Q_5>1$. The natural guess is that while there is an exact CFT for every $Q_1$ and $Q_5$, the NS-NS string theory backgrounds for $Q_5>1$ are given by a Laplace transform over $Q_1$
\begin{equation} \label{eq:bigger_q5_guess}
    Z_{p,Q_5}[J] = \sum_{Q_1=0}^\infty p^{Q_5 Q_1} Z_{Q_1,Q_5}[J],
\end{equation}
such that only $Z_{p,Q_5}$ has a well behaved string perturbation theory. This suggestion is a refinement of the Legendre transform suggested in \cite{Kim:2015gak} (which gives a good approximation for large $Q_1$ or small $g_s$).  Note that when $Q_1$ is relatively prime to $Q_5$ the CFTs appearing here are also on the moduli space of the $N=Q_1 Q_5$ free orbifold as discussed above, while for other values of $Q_1$ they are not; it is not obvious if all values of $Q_1$ should appear in \eqref{eq:bigger_q5_guess} or just the relatively prime ones.

To be precise, for \eqref{eq:bigger_q5_guess} to make sense it is essential to properly normalize the sources in the CFT as a function of $Q_1$ (as we did in \eqref{eq:my_norm} above).
Furthermore, we saw that to get the expected powers of $g_s$, it was necessary to also normalize the sources by a power of $p$ \eqref{eq:op_redef}. This is important if we would like to perform an inverse Laplace transform to fixed $Q_1$. For $Q_5>1$, the operator $I$ (related to $Q_1$) was found to satisfy \cite{Giveon:2001up,Kim:2015gak}
\begin{equation}
    \left\langle I \ \prod_{j=1}^n \Phi_{h_j}(z_j) \right\rangle_{\text{conn},g} = \frac{1}{Q_5}\left(1-g+\sum_{j=1}^n (h_j-1) \right)\cdot \left\langle \prod_{j=1}^n \Phi_{h_j}(z_j) \right\rangle_{\text{conn},g},
\end{equation}
where $h_j$ is the CFT dimension of the operator.
Comparing with \eqref{eq:I_conn}, this suggests that only in the normalization of $\Phi_{h}$ where each string diagram comes with a power of $p^{1-g+\sum_{j=1}^n (h_j-1)}$ does an inverse Laplace transform give the fixed $Q_1$ generating function. 
One can further wonder if something can be said about the appropriate fixed-$Q_1$ normalization. For large $Q_1$, using the relation $Q_1 \propto p$, the two-point function will scale as $\langle \Phi_h(z_1)\Phi_h(z_2)\rangle_{Q_1} \sim Q_1^{2h-1}$.
To find the exact normalization, we need to compute the inverse Laplace transform of the full string theory calculation. 


For $Q_5=1$, since we have an equality in \eqref{eq:grand_sphere_Z} for finite values of $p$, it seems that the perturbative string theory here captures the full partition function of the free symmetric orbifold, with no non-perturbative contributions. 
Naively, for higher genus manifolds, one would expect (based on weakly curved examples) to have different bulk backgrounds that give saddle points leading to non-perturbative contributions (in $g_s$) to the same CFT partition function.
But, as noted in \cite{Eberhardt:2021jvj}, this is not the case here, suggesting that the perturbative expansion around any bulk background (with the appropriate boundary) is complete by itself, giving a strong version of background independence (and in \cite{Eberhardt:2021jvj} this was shown explicitly for some orbifolds of $AdS_3$, see also \cite{Knighton:2024ybs}). 
This is obviously related to the localization of the string worldsheets on the boundary. Presumably, all of these properties will no longer be present once we deform away from the free orbifold theory (and, in particular, for $Q_5 > 1$).

\section{Breakdown of string perturbation theory} \label{sec:breakdown}

\subsection{The large volume limit and S-duality}

As reviewed in section \ref{sec:review}, the symmetric orbifold \eqref{eq:freeorb} is believed to be dual (up to the Laplace transformation reviewed in the previous section) to the NS-NS background with $(Q_1,Q_5)=(N,1)$ and vanishing R-R scalars, where the moduli of the $T^4$ in the symmetric orbifold are identified with the moduli of $T^4$ in string theory (we will discuss the other ${\hat T}^4$ in \eqref{eq:freeorb} in the next section). The six-dimensional string coupling in this background is $g_6^2 = 1/N$ \eqref{eq:ns_ns_desc}, so this description is weakly coupled when $N\gg 1$, and the $1/N$ expansion of the free orbifold correlation functions \cite{Lunin:2000yv,Pakman:2009mi,Eberhardt:2019ywk} was argued to exactly match with the NS-NS perturbative expansion. However, while for some observables the perturbative expansion is governed by $g_6$, for other observables it is governed by $g_{10}$, and, in particular, the latter parameter is expected to control string perturbation theory in the large volume limit. We denote the volume of the torus in string units in this frame, which is also the dual symmetric orbifold metric, by $V$. Since $g_{10}^2 = V/N$ \eqref{eq:ns_ns_desc}, this implies that string perturbation theory should break down when $V$ becomes as large as $N$, even though the orbifold remains free for any value of $V$. Moreover, for $g_{10}^2 \gg 1$ the S-dual R-R background becomes weakly coupled, so we may expect to obtain a different perturbative expansion for the free orbifold in this large-volume regime, that would reproduce this different string perturbation theory. How is this consistent with the free orbifold?

The point is that even though the orbifold \eqref{eq:freeorb} is free for any $V$ and $N$, its (non-trivial) $1/N$ expansion, reviewed in the previous section, can sometimes break down. In particular, it was shown in \cite{Eberhardt:2019ywk} that this expansion can be mapped to the NS-NS perturbative expansion. In this expansion, for large $V \gg 1$, higher genus diagrams are proportional to $V^g$, because this factor would arise (for $V \gg 1$) from the momentum states running in the loops. For instance, the torus partition function of a scalar of radius $R \gg 1$ is proportional to $R$. The general structure of the $1/N$ perturbation theory that arises from the free orbifold (on a sphere) for $V \gg 1$ is thus schematically
\begin{equation} \label{vfactors}
    \langle O_{w_1}(z_1) ... O_{w_n}(z_n) \rangle = N^{1-\frac{n}{2}}\left((g=0) + \frac{V}{N} \cdot (g=1) + \cdots + 
    \left(\frac{V}{N}\right)^{g_\text{max}} \cdot (g=g_\text{max})\right),
\end{equation}
 where $g_\text{max}$ is the maximal genus arising in that correlation function. In the language of the free orbifold, even though it lives on a topologically trivial space, there are non-trivial loops going around branch points where twist operators are inserted, along which the different $T^4$ factors are permuted, and there are non-trivial zero modes of the $T^4$ scalars along these loops, such that the integrals over them give the factors of $V$ in \eqref{vfactors}.
Thus, even though the description \eqref{eq:freeorb} is always free, its $1/N$ expansion indeed breaks down (in any correlation function that contains $g>0$ contributions) once $V$ becomes as large as $N$, consistent with the bulk string theory. 

As mentioned above, S-duality in string theory seems to imply that for $V \gg N$ we should find a new perturbative description of the free orbifold, that would match with the R-R string theory perturbative expansion, that has $g_{10}^2 = N/V$. Indeed, our analysis in section \ref{sec:sugra} tells us that this string theory has a $(Q_1,Q_5)=(N,1)$ R-R description with parameters
\begin{equation} \label{eq:RR_param}
	g_6^2 = \frac{1}{V}, \quad g_{10}^2 = \frac{N}{V}, \quad \frac{R^2}{\alpha'} = \sqrt{\frac{N}{V}}, \quad \frac{\text{Vol}(T^4)}{\alpha^{'2}} = N,
\end{equation}
such that the theory with $V \gg N \gg 1$ seems to be weakly coupled.

While individual correlation functions like \eqref{vfactors} can be rewritten as expansions in $N/V$, it is easy to check that there is no sensible expansion in this parameter of the full theory. The resolution of this apparent paradox is that \eqref{eq:RR_param}, which follows from supergravity, is not valid in this regime, since it gives an AdS radius that is much smaller than the string scale. When this happens, we expect all the stringy modes to be at the AdS energy scale, rather than the string scale. The physics (at least for the gravitational sector) is expected to be governed by Newton's constants in AdS units, which are given by the S-duality invariant values
\begin{equation}\label{eq:newtons2}
     G_N^{(6)}/R^4 = 1/N, \quad G_N^{(10)}/R^8 = V/N.
\end{equation}
The fact that $G_N^{(10)}$ is large in AdS units implies that perturbation theory is not valid, despite the naive expectations from \eqref{eq:RR_param}. The bottom line is, as discussed in \cite{Martinec:2022okx}, that for $Q_5=1$ the R-R picture is always strongly coupled, but the NS-NS picture is weakly coupled for $V \ll N$. For $V \gg N$ both pictures are strongly coupled. It is only for $Q_5 \gg 1$ that there is a region where the R-R description is truly weakly coupled, and both the string couplings and Newton's constants in AdS units are small.

\subsection{Breakdown of perturbation theory at high energies} \label{sec:planck}

There is another regime where we expect string perturbation theory to break down, which is the regime of high energies (scaling as some power of the inverse string scale and the inverse string coupling). In this subsection, we study the $n$-point function of single-cycle operators, all with dimensions scaling as $E\gg 1$, and we will see for which energies the $1/N$ expansion breaks down. Note that in a highly curved background it is not clear from the space-time point of view at which energies this should happen.

If we look at an operator of dimension $L_0=h$ in the $w$'th twisted sector of the orbifold, its energy (the conformal dimension of the corresponding twisted sector operator) is given by
\begin{equation}
    L_0 = \frac{h}{w} + \frac{w-w^{-1}}{4},
\end{equation}
where the second term comes from the dimension of the twist field (in the NS-NS sector of the CFT).
For a given operator with some $h \gg 1$, this implies that the lowest-energy state it gives will come from the sector with $w \simeq \sqrt{h}$ and will have $E = L_0+{\overline L}_0 \simeq \sqrt{h}$. Conversely, this implies that typical operators with energy $E$ will come from sectors with $w \simeq E$.

We start by estimating the number of diagrams with genus $g$ that contribute to the $n$-point correlation function, following \cite{Gaberdiel:2020ycd}.
Each genus $g$ diagram satisfies
\begin{equation}
    n - n_E + n_F = 2-2g,
\end{equation}
with $n_E, n_F$ the number of edges and faces of the skeleton graph. 
For high enough $E$, we can approximate the graphs being triangular $3 n_F = 2 n_E$, which gives
\begin{equation}\label{eq:n_E}
    n_E = 3(2g-2+n).
\end{equation}
The diagram is completely determined by the number of edges between each pair of vertices $i$ and $j$, which we denote by $n_{ij}$. In a graph with $n_E$ edges,  $n_{ij}$ has $n_E$ \eqref{eq:n_E} non-trivial elements, and it has to satisfy $\sum_{j} n_{ij} = w_i$, giving $n$ constraints. So overall there are 
$3(2g-2)+2n$ independent variables $n_{ij}$, each of order $\sim w \sim E$. In this way we can approximate the number of diagrams by
\begin{equation}\label{eq:number_of_covers}
    \# (\text{coverings with genus }g) \simeq E^{3(2g-2)+2n}.
\end{equation}

It is not clear how the contribution of individual diagrams scales in the large energy / cycle length limit. If we assume that the contribution of each diagram separately does not scale with the energy, then the $1/N$ expansion breaks down when the number of diagrams grows faster than $N^g$, which happens (using \eqref{eq:number_of_covers}) when $E \sim N^{1/6}$. If individual diagrams grow with the energy in a way that depends on the genus, the expansion could break down at lower scales, such as the naive ten dimensional Planck scale coming from \eqref{eq:newtons_V}, which scales as $N^{1/8}$. In any case, we find that perturbation theory breaks down at high energies going as a negative power of the string coupling, as expected.

\section{The decoupled sector} \label{sec:decoupled_sec}
\subsection{The decoupled sector of the \titlemath{$Q_5=1$} CFT} \label{sec:q5_1_dec_sector}

In section \ref{sec:dual_cft} we reviewed the argument that the SCFT dual of the $Q_5=1$, $Q_1=N$ NS-NS background with vanishing R-R scalars is given by a product of the symmetric orbifold over $T^4$ and an extra $\hat T^4$ \eqref{eq:freeorb}. The symmetric orbifold ${\rm Sym}_N(T^4)$ has 4 holomorphic and 4 anti-holomorphic $U(1)$ currents associated with the ``center-of-mass'' momentum and winding of the $T^4$, whose currents are given by the sums of the corresponding $U(1)$ currents over the $N$ copies of the $T^4$, such that they have level $N$.
The $\hat T^4$ CFT has its own $8$ (winding and momentum) currents, all at level $1$. 

Invoking the coset construction, it is possible to rewrite this theory as the semi-direct product
\begin{equation}\label{eq:cft_coset}
    (T^{4})^N/S_N\times \hat T^4 \equiv 
    \left((T^{4})^N/S_N\right)/U(1)_N^8 \rtimes \left(U(1)_N^8 \times \hat U(1)_1^8\right).
\end{equation}
We will name the first term on the right-hand side the ``coset CFT'', and the second the ``Sugawara CFT''.
The operators of the full theory are spanned by products of operator pairs, one from the coset and one in the Sugawara $U(1)_N^8 \times \hat U(1)_1^8$ theory; the product with $U(1)_N^8$ is semi-direct, meaning that not all possible products of operators appear in the theory, while the product with ${\hat U(1)}_1^8$ is direct, such that the theory contains any operator in this sector multiplied by any operator in the rest of the CFT.

The left-moving and right-moving energies of the full theory can each be written as a sum
\begin{equation}
    E^{L/R} = E^{L/R}_\text{coset} + E^{L/R}_\text{Sugawara},
\end{equation}
with $E_\text{Sugawara}$ given by the Sugawara stress tensor of the $8$ left-moving or right-moving currents. Due to unitarity, the Sugawara energy of a state gives a lower bound for the left and right moving energies
\begin{equation}
    E^{L/R} \ge E^{L/R}_\text{Sugawara}.
\end{equation}

We denote the metric and $B$-field of the $T^4$ in the orbifold by $E = G+B$, and $\hat E = \hat G + \hat B$ for the decoupled $\hat T^4$. We also denote the integer winding and momentum charges of the orbifold by $w^i$, $p_i$, and the $\hat T^4$ charges by $\hat w^i$, $\hat p_i$ ($i=1,\cdots,4$). In this notation, the left/right Sugawara energy is given by
\begin{equation}\label{eq:sugawara_bound_can}
\begin{split}
    E_\text{Sugawara}^{L/R} &=\frac{1}{4}
    \begin{bmatrix}
        w & p & \hat w & \hat p
    \end{bmatrix}
    \begin{bmatrix}
        \frac{1}{N}(G+B^T G^{-1} B) & \frac{1}{N}(\pm 1 + B G^{-1}) & 0 & 0\\
        \frac{1}{N}(\pm 1 - G^{-1} B) & \frac{1}{N} G^{-1} & 0 & 0\\
        0 & 0 & \hat G + \hat B^T \hat G^{-1} \hat B & \pm 1 + \hat B \hat G^{-1}\\
        0 & 0 & \pm 1 -\hat G^{-1} \hat B & \hat G^{-1}
    \end{bmatrix}
    \begin{bmatrix}
        w\\
        p \\
        \hat w\\
        \hat p
    \end{bmatrix}.
\end{split}
\end{equation}

We now turn to the string theory background with NS-NS fluxes\footnote{In this section we ignore the fact that perturbative string theory in this background actually involves a grand-canonical ensemble of CFTs, as discussed in section \ref{sec:st_grand_can}; namely, we consider the theory with fixed $Q_1$ and $Q_5$, even though its perturbative string description is more complicated. The U-duality group that we discuss in this section acts naturally on the backgrounds with fixed $Q_1$ and $Q_5$, rather than on the grand-canonical ensemble that appears in the perturbative NS-NS-background string. }. The charges $w^i$ and $p_i$ correspond to fundamental string winding and momentum on the $T^4$, respectively. The extra $\hat T^4$ accounts for the charges of R-R D-branes wrapping on the $T^4$. In type IIB string theory, we have 4 charges from D1-branes winding on one of the $T^4$ circles $w^i_\text{D1}$, and $\binom{4}{3}=4$ from D3-branes winding on $3$ circles of the $T^4$, $w^\text{D3}_{i} =\epsilon_{ijkl} w_\text{D3}^{jkl}$. We identify the $\hat T^4$ charges with the bulk R-R charges by
\begin{equation}\label{eq:winding_dict}
    \hat w^i = w^i_\text{D1}, \quad \hat p_i = w^\text{D3}_{i}.
\end{equation}

What is the meaning of the Sugawara energy bound from the string theory side? This question was answered in \cite{Larsen:1999uk}. The authors used the description of the $AdS_3 \times S^3 \times T^4$ background as a near-horizon limit of a flat space brane configuration. The flat-space BPS formula holds even in the decoupling limit, and when subtracting the energies of the strings and 5-branes making up the background, it gives an energy bound\footnote{This is obvious in the R-R sector of the CFT, in which the fermions are periodic around the circle and supersymmetry is preserved. Here we discuss the NS-NS sector of the CFT which is dual to string theory on $AdS_3$ (where the spatial circle of the CFT is contractible), where the fermions are anti-periodic, so that the configuration does not directly arise as a near-horizon limit. However, ${\cal N}=(4,4)$ supersymmetry implies that this sector is related by spectral flow to the R-R sector, so the same BPS bounds hold.} on particle-like excitations in $AdS_3$. We will discuss the energy formula for a general background in the next subsection. For the $(Q_1,Q_5)=(N,1)$ NS-NS background with $T^4$ metric $G$ and vanishing $B$-field and R-R scalars, the resulting BPS formula is \cite{Larsen:1999uk}
\begin{equation}\label{eq:bps_energy_canonical}
\begin{split}
    E_\text{BPS}^{L/R}
    &= 
    \frac{1}{4 N^{1/2}} \left(g_{10} w^2 + g_{10}^{-1} w_{D1}^2 
    + (g_{10}/V) \ p^2 + (V/g_{10}) \ (w^{D3})^2 \right)
    \pm \frac{1}{2N} p \cdot  w \pm \frac{1}{2} w_{D1} \cdot w^{D3} \\
    &= \frac{1}{4N}\left(V^{1/2} w^2 + V^{-1/2} p^2 \pm 2 w \cdot p\right)
    +\frac{1}{4}\left(V^{-1/2} \hat w^2 + V^{1/2} \hat p^2 \pm 2 \hat w \cdot \hat p \right).
\end{split}
\end{equation}
All the (implicit) $T^4$ metric contractions in this equation are written in terms of the unit-volume metric $(G_0)_{ij} = V^{-1/2} G_{ij}$, where $V$ is the $T^4$ volume $V = \sqrt{\det(G)}$. In the second line we used \eqref{eq:winding_dict} and the relation \eqref{eq:ns_ns_desc}
\begin{equation}
    V = N g_{10}^{2}.
\end{equation}
To find the relation to the CFT parameters, we compare the bulk BPS bound \eqref{eq:bps_energy_canonical} to the CFT Sugawara bound \eqref{eq:sugawara_bound_can}. The orbifold metric is immediately identified with the bulk $T^4$ metric, while the $\hat T^4$ metric has the same shape but an inverse volume, $\hat G = \frac{1}{V} G$. The fact that we got an inverse volume can be traced to the fact that the D1-brane tension is proportional to $g_{10}^{-1} \sim V^{-1/2}$ (or, alternatively, to the fact that this extra ${\hat T}^4$ comes from Wilson lines which live on a dual torus). Choosing a different $\hat G$ in the CFT corresponds in the bulk to different boundary conditions for the $(U(1)\times U(1))^4$ R-R gauge fields (analogous to a double-trace deformation).

We can generalize \eqref{eq:bps_energy_canonical} to non-zero bulk B-fields, by performing the transformation \cite{Larsen:1999uk}
\begin{equation}\label{eq:add_B}
    p_i \mapsto p_i + B_{ij} w^j, \quad
    \hat w^i \mapsto \hat w^i - \frac{1}{2} \epsilon^{iljk} B_{jk} \hat p_{l}.
\end{equation}
The first transformation gives a $B$-field (identical to the one in the bulk) to the $T^4$ in the orbifold CFT. The second (coming from the $C_2 \wedge B_2$ coupling in the D3-brane action) amounts to adding $B^{' ij} = -\frac{1}{2} \epsilon^{ijkl} B_{kl}$ in the T-dual frame of the $\hat T^4$ sigma-model. Altogether, the CFT moduli are related to the string theory moduli (for vanishing R-R scalars) by
\begin{equation}\label{eq:bulk_to_cft_GB}
    E_{ij} = G_{ij} + B_{ij}, \quad
    (\hat E)^{-1, ij} = V \cdot G^{-1, ij} - \frac{1}{2} \epsilon^{ijkl} B_{kl}.
\end{equation}

What is the worldsheet interpretation of the decoupled sector? The momentum and winding currents of the orbifold have a clear worldsheet dual, coming from the metric and B-field of the bulk $T^4$, whose $U(1)^8$ currents can be used to construct the corresponding currents of the CFT (for $Q_5>1$ this was discussed in \cite{Kutasov:1999xu,Gaberdiel:2011vf}, and for $Q_5=1$ in \cite{Gaberdiel:2021njm}). Perturbative string states are charged under these currents.
On the other hand, the charged states under the ${\hat U(1)}^8_1$ R-R currents can all be constructed as ``boundary modes'' of the bulk Chern-Simons theory (for general level $k$ this is only true for states whose charges are integer multiples of $k$, while here $k=1$). In the bulk they are
pure-gauge modes of the R-R gauge potentials. One would naively expect the $\hat T^4$ currents to have their own worldsheet vertex operator, describing the R-R potential in the bulk\footnote{One reason why this is not obvious is that in flat space, vertex operators for R-R potentials, as opposed to field strengths, exist in some pictures for the worldsheet ghosts but not in others.}, and to have D-branes (though no perturbative string states) that carry their charge.
However, the structure of the dual CFT contradicts this expectation.\footnote{We thank Sameer Murthy and Mukund Rangamani for discussions on this issue.} Notice that each single-string state has its single-string descendants under the NS-NS $U(1)^8_N$ current algebra. In the CFT these are descendants in the core CFT whose symmetric orbifold we are considering. 
More generally, the existence of a worldsheet vertex for a CFT current means there's a $U(1)^8$ current algebra module already in the single-string partition function $Z_\text{1-string} = \log (Z)$. However, the partition function of the $\hat T^4$ CFT is simply an overall factor in the full partition function. In terms of the string theory, for every multi-string state, each of its single-strings has its own NS-NS $U(1)^8_N$ descendants, but only one multi-string R-R ${\hat U(1)}^8_1$, or $\hat T^4$, descendant. This suggests that the $\hat T^4$ ${\hat U}(1)^8_1$ currents have no local string theory vertex.\footnote{It is possible that worldsheet operators for the total charge of these currents can be defined.}

An alternative perspective on this is that a $(U(1)\times U(1))^4_1$ Chern-Simons theory on $AdS_3$ is dual by itself to a ${\hat T}^4$ CFT, with the moduli of the ${\hat T}^4$ determined by the boundary conditions on the CS fields (see, for instance, \cite{Gukov:2004id,Aharony:2023zit}). So this sector of the theory, which is modular invariant by itself, decouples completely from the rest of the string theory; it has no interactions with any string states (and does not even depend on the topology of the interior of the AdS space, just on the boundary conditions). In particular, even though naively one may expect the string theory with $Q_5=1$ to contain particle-like D-branes wrapped on cycles of the $T^4$ which are charged under these ${\hat U(1)}$'s, such D-branes have not yet been found, and this analysis implies that they should not exist as non-trivial boundary states.

This behavior is special to level $k=1$, and does not apply to the decoupled $U(1)^8_N$ sector; we will discuss in what sense it is decoupled later, after generalizing to other values of $Q_5$.

\subsection{String dualities and exact CFT deformations}

We now extend our discussion to more general string theory backgrounds with different charges, following \cite{Larsen:1999uk}. We write the string and 5-brane integer fluxes as a charge matrix 
\begin{equation} \label{onefivecharges}
    Q = \begin{bmatrix}
    f_1 & d_1\\
    -d_5 & n_5
    \end{bmatrix},
\end{equation}
which labels the (integer) number of F1, D1, D5, and NS5-branes in the original flat-space construction (we assume for simplicity there are no wrapped D3-branes). Before taking the near-horizon limit, the backgrounds are also labeled by $\tau = \chi + i g_{10}^{-1}$ and by $\tilde \tau = A_4 +i v_4 g_{10}^{-1}$. Here $g_{10}$ is the ten-dimensional string coupling, $v_4=\sqrt{{\rm det}(G)}$ the volume of the $T^4$, and $\chi$ and $A_4$ are the R-R scalar $C_0$ and the holonomy of the R-R 4-form $C_4$ on the $T^4$, respectively.\footnote{We assume here vanishing $B_2$ and $C_2$ fields on the $T^4$, although the generalization is straightforward \cite{Larsen:1999uk}.} It is useful to write $\tau,\tilde\tau$ also as matrices
\begin{equation}\label{eq:T_def}
    \mathcal{T} = g_s \begin{bmatrix}
        1 & -\chi \\
        -\chi & g_s^{-2} + \chi^2
    \end{bmatrix}, \quad
    \tilde{\mathcal{T}} = \frac{g_s}{v_4} \begin{bmatrix}
        1 & -A_4 \\
        -A_4 & v_4^2/g_s^2 + A_4^2
    \end{bmatrix}.
\end{equation}
When taking the near-horizon limit of a brane configuration with a charge matrix $Q$, the attractor mechanism gives a relation between $\tau$ and $\tilde\tau$ of the form
\begin{equation}\label{eq:min_cond}
    \det(Q) \cdot 1 = \tilde{\mathcal{T}} Q \mathcal{T} Q^T.
\end{equation}
This minimizes the tension formula $\text{tr}(\tilde{\mathcal{T}} Q \mathcal{T} Q^T)$. 
For example, in the case of the $d_1,d_5$ background the equation is $\tilde \tau = (d_1/d_5) \tau$.

As in the $Q_5=1$ case, we now consider particle excitations in this background. These are labeled by their $T^4$ winding and momentum $w^i_\text{F1}$, ${\tilde p}_i$, and by the R-R winding charges $w_\text{D1}^i$ and $w^\text{D3}_i$. It is useful to organize them in pairs based on their natural $T^4$ indices
\begin{equation}
    q^i = \begin{bmatrix}
        w_{F1}^i\\
        w_{D1}^i
    \end{bmatrix}, \quad
    q_i = \begin{bmatrix}
        {\tilde p}_i\\
        w^{D3}_i
    \end{bmatrix} 
    .
\end{equation}
For a general string theory background, the energy bound found by \cite{Larsen:1999uk} is given by
\begin{equation}\label{eq:covariant_energy_formula}
    E_\text{BPS}^{L/R} = \frac{1}{4} \begin{bmatrix}
        q^i & q_j
    \end{bmatrix}
    \begin{bmatrix}
        \frac{\mathcal{T}}{\det^{1/2}(Q)} & \pm Q^{-1}\\
        \pm (Q^{-1})^T & \frac{\tilde{\mathcal{T}}}{\det^{1/2}(Q)}
    \end{bmatrix}
    \begin{bmatrix}
        q^k \\
        q_l
    \end{bmatrix},
\end{equation}
where any $T^4$ index contraction is written in terms of the unit metric $v_4^{-1/2} G$ (the $v_4$ dependence appears through $\mathcal{T}, \tilde{\mathcal{T}}$).

The U-duality group of the system is $SO(5,5;\mathbb{Z})$. In this section we are interested only in the $SL(2,\mathbb{Z})_L\times SL(2,\mathbb{Z})_R$ U-duality subgroup which fixes the general form \eqref{onefivecharges} of the charges, the vanishing of $B_2$ and $C_2$, and the unit-volume metric on the $T^4$. $SL(2,\mathbb{Z})_R$ is generated by the S-duality of type IIB string theory and by the transformation $\chi \mapsto \chi+1$. $SL(2,\mathbb{Z})_L$ is generated by $T_{1234} S T_{1234}$ (where $T_{1234}$ means a T-duality transformation on the four cycles of the $T^4$) and by the transformation $A_4 \mapsto A_4+1$.
For $g_L,g_R \in SL(2,\mathbb{Z})$, the background parameters are transformed by
\begin{equation}\label{eq:QT_trans}
\begin{split}
    Q' = g_L Q g_R^{T}, \quad \mathcal{T}' = g_R^{-1,T} \mathcal{T} g_R^{-1}, 
    \quad \tilde{\mathcal{T}}' = g_L^{-1,T} \tilde{\mathcal{T}} g_L^{-1}.
\end{split}
\end{equation}
The particle charges are transformed by
\begin{equation}\label{eq:q_trans}
    q^{i\prime} = g_R \cdot q^i, \quad q_j' = g_L \cdot q_j.
\end{equation}
We note that the energy formula \eqref{eq:covariant_energy_formula} (as well as the relation \eqref{eq:min_cond}) is duality-invariant under the transformation of the background parameters and the charge lattice.


We would like to describe the structure of the conformal manifold for \eqref{eq:cft_coset}. Different values of $\tau,\tilde\tau$ correspond to different deformations of \eqref{eq:cft_coset}, but because $\tau$ and $\tilde \tau$ are related through \eqref{eq:min_cond}, each deformation can be labeled by $\tau$ alone. 
The $(N,1)$ NS-NS background is given by
\begin{equation}\label{eq:canonical}
    Q = \begin{bmatrix}
    N & 0\\
    0 & 1
    \end{bmatrix},
\end{equation}
and we consider values of the charges \eqref{onefivecharges} that can be mapped to these charges by a U-duality transformation. We can then describe the deformations by $\tau$ in this background, and
two deformations are identical if and only if there is a U-duality transformation mapping one $\tau$ to the other, while keeping the canonical background \eqref{eq:canonical} fixed. In other words, the conformal manifold is given by the fundamental domain of the U-duality subgroup which stabilizes the canonical background \eqref{eq:canonical}.\footnote{Of course, we could just as well describe the geometry of the conformal manifold in terms of deformations of any other string theory background U-dual to \eqref{eq:canonical}.} As explained in \cite{Larsen:1999uk}, this subgroup is isomorphic to the congruence modular subgroup $\Gamma_0(N)$, or the set of pairs
\begin{equation}\label{eq:Gamma0}
    g_L =  \begin{bmatrix}
    \alpha & \beta N \\
    \gamma & \delta
    \end{bmatrix}, \quad
    g_R = \begin{bmatrix}
    -\delta & -\gamma N\\
    -\beta & -\alpha
    \end{bmatrix},
\end{equation}
with $\alpha \delta - \beta \gamma N=1$. 
Figure \ref{fig:N30_moduli} depicts the fundamental domain of $\Gamma_0(N)$ for $N=30$, in terms of the S-dual parameter $\tau_{RR} = -1/\tau$.
The grey dashed lines separate the domain into (halves) of $SL(2,\mathbb{Z})$ fundamental domains. The number of $SL(2,\mathbb{Z})$ fundamental domains in the fundamental domain of $\Gamma_0(N)$ is given by \cite{Larsen:1999uk,schoeneberg2012elliptic}
\begin{equation}
    (SL(2,\mathbb{Z}):\Gamma_0(N)) = N \prod_{\substack{p\mid N\\ p \text{ is prime}}} (1+p^{-1}).
\end{equation}
The vertical black line corresponds to the $(N,1)$ background with vanishing R-R scalars $\chi=A_4=0$. The point $\tau_\text{RR} = 0$ corresponds to the $g_{10}=0$ limit of the NS-NS background. The point $\tau_\text{RR} = i \infty$ is the strongly coupled limit, or, by S-duality, the $g_{10}=0$ limit of the $(N,1)$ R-R background.

Every NS-NS (and R-R) $(Q_1,Q_5)$ background with $Q_1$,$Q_5$ mutually prime is related to the canonical $(N,1)$ background \eqref{eq:canonical} with $N=Q_1\cdot Q_5$ by a U-duality transformation, and
can be described as a CFT deformation of \eqref{eq:cft_coset}. 
We are specifically interested in the singular locus of these backgrounds (with vanishing R-R scalars $\chi=A_4=0$), given by
\begin{equation}\label{eq:q1q5_background}
    Q = \begin{bmatrix}
    Q_1 & 0\\
    0 & Q_5
    \end{bmatrix}, \quad 
    \mathcal{T} = \begin{bmatrix}
        t & 0 \\
        0 & t^{-1}
    \end{bmatrix}, \quad 
    \tilde{\mathcal{T}} = \begin{bmatrix}
        t^{-1} Q_5/Q_1 & 0 \\
        0 & t Q_1/Q_5
    \end{bmatrix},
\end{equation}
with $t$ the ten-dimensional string coupling in the NS-NS $(Q_1,Q_5)$ background.
These backgrounds are related to the canonical background \eqref{eq:canonical} by the duality transformation 
\begin{equation}\label{eq:ns_gs}
    g_L =  \begin{bmatrix}
    a Q_5 & b Q_1 \\
    1 & 1
    \end{bmatrix}, \quad
    g_R = \begin{bmatrix}
    Q_5 & -Q_1\\
    -b & a
    \end{bmatrix},
\end{equation}
with $a$, $b$ integers satisfying $a Q_5 - b Q_1 = 1$. Under the transformation \eqref{eq:QT_trans}, the dual canonical parameters are
\begin{equation} \label{eq:q1q5_can_var}
    g_{10} = a^2 t + b^2 t^{-1}, \quad 
    \chi = -\frac{Q_1 a t + Q_5 b t^{-1}}{a^2 t + b^2 t^{-1}}.
\end{equation}

Written in terms of $\tau_\text{RR} = - (\chi + i/g_{10})^{-1}$, these lines are drawn as thick black arcs in figure \ref{fig:N30_moduli}. Each arc is drawn twice, related by a sign $\chi\mapsto -\chi$, but the two arcs are identified by the action of $\Gamma_0(N)$.
The limit $t=0$ is the weakly coupled limit of the NS-NS $(Q_1,Q_5)$ background, which is also dual by a $T_{1234}ST_{1234}$ duality transformation to the weakly coupled R-R $(Q_5,Q_1)$ background.\footnote{By weak coupling here we mean a small ten-dimensional string coupling; as discussed above, this does not necessarily imply that the six-dimensional string coupling or Newton's constant are small, so the theory may not really be weakly coupled.} Both are mapped to the point $\tau_\text{RR} = \pm b/Q_5$ in the figure. The limit $t=\infty$ in \eqref{eq:q1q5_background} corresponds by S-duality to the weakly coupled R-R $(Q_1,Q_5)$ background, and by another $T_{1234}ST_{1234}$, to the weakly coupled NS-NS $(Q_5,Q_1)$ background. Both are mapped to $\tau_\text{RR} = \pm a/Q_1$ in the figure.
Generally, the arcs of the fundamental domain's boundary are identified in a complicated way. 
We conjecture that the CP-invariant line $\chi = \pm 1/2$ in the figure (in the R-R background) corresponds to the orbifold \eqref{eq:freeorb} at the singular point $\theta=0$ for the 2-cycle at the $\mathbb{Z}_2$ fixed point; note that this line has $A_4 = \pm N/2$ both in the R-R and in the NS-NS descriptions\footnote{The line $\chi=\pm N/2$ in the $(N,1)$ NS-NS background, which is related to the $A_4 = \pm N/2$ line by T-duality in the NS-NS frame, and that appears
in the figure as the two (identified) boundary arcs emanating from $\tau_\text{RR}=0$, is also singular.}.

For a given $t$, the string theory background \eqref{eq:q1q5_background} is a deformation of the CFT \eqref{eq:cft_coset}; this deformation involves both a blow-up of the $\mathbb{Z}_2$ singularity of the orbifold, which affects only the ``coset CFT'', and a $J {\bar J}$ deformation, which affects only the ``Sugawara CFT''. In the dual $(N,1)$ background, it is described by $\tau$ and $\tilde \tau$ given by \eqref{eq:q1q5_can_var}. Consider a general deformation labeled by $\tau,\tilde\tau$. For the canonical background \eqref{eq:min_cond} the two are related by 
\begin{equation}\label{eq:can_min_cond}
    A_4 = -v_4 \chi, \quad v_4 = \frac{V}{1+ \chi^2 \frac{V}{N}},
\end{equation}
with $V = N g_{10}^2$. Labelling a general deformation by $V, \chi$, the BPS energy bound \eqref{eq:covariant_energy_formula} gives in the canonical frame (assuming $B=0$)
\begin{equation}\label{eq:bps_chi}
\begin{split}
    E_\text{BPS}^{L/R} &= \frac{1}{4} \begin{bmatrix}
        w & p & \hat w & \hat p
    \end{bmatrix}
    \begin{bmatrix}
        \frac{V^{1/2}}{N} & \pm \frac{1}{N} & - \chi \frac{V^{1/2}}{N} & 0\\
        \pm \frac{1}{N} & \frac{1+ \frac{V}{N} \chi^2}{N V^{1/2}} & 0 & \chi \frac{V^{1/2}}{N}\\
        - \chi \frac{V^{1/2}}{N} & 0 & \frac{1+ \frac{V}{N} \chi^2}{V^{1/2}} & \pm 1\\
        0 & \chi \frac{V^{1/2}}{N} & \pm 1 & V^{1/2}
    \end{bmatrix}
    \begin{bmatrix}
        w \\
        p\\
        \hat w\\
        \hat p
    \end{bmatrix}\\
    & = \frac{1}{4N}\left(V^{1/2} (w- \chi \hat w)^2 + V^{-1/2} p^2 \pm  2 w \cdot p \right)\\
    &\quad + \frac{1}{4}\left(V^{-1/2} \hat w^2 + V^{1/2} \left(\hat p+\frac{\chi}{N} p\right)^2 \pm  2 \hat w \cdot \hat p\right).
\end{split}
\end{equation}
This generalizes \eqref{eq:bps_energy_canonical} to the case of non-zero $\chi$. We see that in the BPS formula the $\chi$ deformation is equivalent to the shift
\begin{equation}
    w^i \mapsto w^i - \chi \hat w^i, \quad \hat p_j \mapsto \hat p_j +\frac{\chi}{N} p_j.
\end{equation}

We would like to repeat the exercise of the previous section. Interpreting the BPS bound \eqref{eq:bps_chi} as the Sugawara energy, we can find the Sugawara CFT deformation for every $V, \chi$. However, unlike \eqref{eq:bps_energy_canonical}, \eqref{eq:bps_chi} is no longer decoupled between the charges of $T^4$ and $\hat T^4$. It is therefore not enough to simply deform the $G$, $\hat G$ metrics. As explained in \cite{Larsen:1999uk}, the Sugawara CFT needs to be further deformed by a mixed $J{\bar J}$ deformation to match with \eqref{eq:bps_chi}.

Let us denote the $U(1)^8_N$ orbifold currents by $J^i$, $\bar J^i$, and the $U(1)^8_1$ $\hat T^4$ currents by $\hat J^i$ and $\hat {\bar J}^i$ ($i=1,\cdots,4$). All the currents are normalized such that the corresponding charges are integers (see appendix \ref{app:orbifold}). A general $J{\bar J}$ marginal deformation is of the form
\begin{equation} \label{eq:JJ_def}
    \Delta S  = 2\pi 
    \int d^2 z \left(
    M_{ij}^{11} J^i \bar J^j + 
    M_{ij}^{12} J^i \hat {\bar J}^j + 
    M_{ij}^{21} \hat J^i \bar J^j + 
    M_{ij}^{22} \hat J^i \hat {\bar J}^j
    \right)
    ,
\end{equation}
for some $8\times 8$ matrix $M_{ij}^{\alpha\beta}$ ($\alpha,\beta = 1,2$, $i,j = 1,\cdots,4$). 
The $M^{22}$ matrix simply deforms the $\hat T^4$ moduli, $\hat E\mapsto \hat E + M^{22}$.
$M^{11}, M^{12}$ and $M^{21}$ don't have such an interpretation, but together with $M^{22}$ they label an $SO(8,8)/(O(8)\times O(8))$ conformal manifold. In the appropriate sense (which takes care of the different levels), $M_{ij}^{\alpha\beta}$ deforms the metric and the $B$-field on the $T^4\times \hat T^4$ manifold.

For every $V$ and $\chi$, it is possible to find the exact $M_{ij}^{\alpha\beta}$ deformation such that the new $E_\text{Sugawara}$ will exactly agree with the string theory BPS bound \eqref{eq:bps_chi}. Starting from the moduli \eqref{eq:bulk_to_cft_GB} with the appropriate $V$, the answer is
\begin{equation}\label{eq:JJ_deformation}
\begin{split}
    M^{11}_{ij} &= 0,\\
    M^{12}_{ij} &= M^{21}_{ij} = - \frac{\chi}{N} G_{ij},\\
    M^{22}_{ij} &= \frac{V}{N} \chi^2 \hat G_{ij} = 
    \frac{\chi^2}{N} G_{ij},
\end{split}
\end{equation}
where in the last equation we used $\hat G = \frac{1}{V} G$. Specifically, plugging in \eqref{eq:q1q5_can_var} will give us the exact Sugawara CFT deformation which corresponds to the singular $(Q_1,Q_5)$ string theory backgrounds.

Let us emphasize that the deformation labeled by $\chi$ on the string theory side deforms both the coset CFT and the Sugawara CFT in \eqref{eq:cft_coset}. The coset deformation is known to correspond at leading order to one of the blow-up modes of the $\mathbb{Z}_2$ orbifold singularity \cite{David:2002wn}. Following the deformation is a highly complicated task, even in conformal perturbation theory, and it is beyond the scope of this paper (see \cite{Benjamin:2021zkn,Guo:2022ifr,Apolo:2022fya,Guo:2022zpn,Fiset:2022erp,Hughes:2023apl,Hughes:2023fot,Gaberdiel:2023lco,Frolov:2023pjw} and references therein for recent progress on this). In this section we identified the exact deformation only of the Sugawara part of the CFT, not the coset CFT. This was possible because the $J{\bar J}$ deformation is integrable from the CFT perspective, and can be matched exactly to the string theory using string dualities.

\subsection{The decoupled sector of the \titlemath{$Q_5>1$} CFT}

In section \ref{sec:q5_1_dec_sector} we asked for the worldsheet interpretation of the decoupled $\hat T^4$ currents. For the backgrounds \eqref{eq:q1q5_background} at small $g_{10}=t$ there is a different perturbative worldsheet description with $Q_5 > 1$ \cite{Giveon:1998ns,Kutasov:1999xu,Giveon:2001up,Maldacena:2000hw}. We would like to ask more generally, what is the $Q_5>1$ worldsheet interpretation of the deformed Sugawara CFT \eqref{eq:JJ_deformation}? 

First, it is useful to re-write the Sugawara energy \eqref{eq:bps_chi} in terms of the $(Q_1,Q_5)$ charges, and not the original CFT charges. The two charge lattices are related through \eqref{eq:q_trans}, \eqref{eq:ns_gs} by
\begin{equation}\label{eq:charge_rel_inv_1}
    \begin{bmatrix}
        w^i_{F1}\\
        w^i_{D1}
    \end{bmatrix}
    = \begin{bmatrix}
        a & Q_1\\
        b & Q_5
    \end{bmatrix}
    \begin{bmatrix}
        w^i\\
        \hat w^i
    \end{bmatrix}, \quad
    \begin{bmatrix}
        {\tilde p}_j\\
        w_j^{D3}
    \end{bmatrix} = 
    \begin{bmatrix}
        1 & -b Q_1 \\
        -1 & a Q_5
    \end{bmatrix} 
    \begin{bmatrix}
        p_j\\
        \hat p_j
    \end{bmatrix}.
\end{equation}
In terms of the $(Q_1,Q_5)$ string theory charges, the Sugawara energy (also given directly from \eqref{eq:covariant_energy_formula}) takes the simple form
\begin{equation}\label{eq:E_ns_sug}
\begin{split}
    E_\text{Sugawara}^{L/R} &= \frac{1}{4 Q_1} \left(
    v_4^{1/2} w_\text{F1}^2 + v_4^{-1/2} {\tilde p}^2 \pm 2 w_\text{F1}\cdot {\tilde p}\right)\\
    &+ \frac{1}{4 Q_5} \left(
    v_4^{1/2} (w^\text{D3})^2 + v_4^{-1/2} w_\text{D1}^2 \pm 2 w_\text{D1}\cdot w^\text{D3}\right),
\end{split}
\end{equation}
where $v_4 = t^2 Q_1/Q_5$ is the worldsheet $T^4$ volume \eqref{eq:ns_ns_desc}. Equation \eqref{eq:E_ns_sug} should be understood as the generalization of \eqref{eq:bps_energy_canonical} for $Q_5>1$. It agrees with the fact that, as reviewed in section \ref{sec:review}, the bulk theory includes a $U(1)^8_{Q_1}\times U(1)^8_{Q_5}$ CS theory in $AdS_3$, from the $4+4$ NS-NS and $4+4$ R-R bulk gauge fields, respectively. The $J{\bar J}$ deformation at these points in the moduli space has precisely the effect of forming orthogonal linear combinations of the original currents of \eqref{eq:cft_coset} that have these new levels. Each set of currents involves linear combinations of the currents from the orbifold and from ${\hat T}^4$.

Just as in the $Q_5=1$ case, the NS-NS currents at level $Q_1$ have a string worldsheet vertex operator, found explicitly in \cite{Kutasov:1999xu}. As the authors comment there, the R-R currents do not seem to appear in the string spectrum, consistent with the fact discussed above that single-string states should have descendants under $8$ of the $U(1)$ symmetries but not under the other $8$.\footnote{Note that the higher Kaluza-Klein modes of the R-R potential which have a non-zero R-R field strength do have a worldsheet vertex operator \cite{Kutasov:1999xu}, but here we consider the pure-gauge mode of the R-R potential.} This also agrees with the string 1-loop calculation done recently for $Q_5>1$ \cite{Mukund:2024}, in which the $U(1)^8_{Q_5}$ R-R currents were absent.\footnote{We thank Rohit Kalloor for his help in clarifying this issue.}

After discussing the currents of the decoupled Sugawara sector, we would next like to understand the bulk interpretation of its charged states. Consider first the undeformed theory \eqref{eq:cft_coset} for $Q_5=1$. A general charged state will satisfy the Sugawara/BPS bound, but will not saturate it. The operators that saturate the bound are exactly the operators that are given entirely by the Sugawara CFT, with only the identity in the coset CFT. These operators can have any $\hat w$, $\hat p$ integer charges (any operator from $\hat T^4$), but the charges $w,p$ must be integer multiples of $N$. We term this charge sub-lattice the BPS lattice\footnote{Note that the states saturating the Sugawara/BPS bound are BPS states before taking the near-horizon limit, where the supersymmetry algebra has central charges corresponding to the momenta and winding numbers on the $T^4$. These states are not BPS states of the theory on $AdS_3$, where there are different BPS states, carrying charges under the $SU(2)_L\times SU(2)_R$ symmetry; we do not discuss these states here. The bound on the energy of the states we discuss in $AdS_3$ comes from the form of the CFT energy-momentum tensor discussed above, rather than from the supersymmetry algebra.}.
We can immediately generalize this statement to the $(Q_1, Q_5)$ backgrounds. Under the $\chi$, $V$ deformation, the energy of the coset changes in a non-trivial way. However, for the BPS lattice, the coset operator is simply the identity. Therefore, also at finite $\chi$ these operators remain BPS, as their energy is given exactly by the (deformed) Sugawara energy \eqref{eq:E_ns_sug}. In order to understand what are these operators from the perspective of the $(Q_1, Q_5)$ background, we rewrite \eqref{eq:charge_rel_inv_1} as
\begin{equation}\label{eq:charge_rel_inv_2}
    \begin{bmatrix}
        \frac{w^i_{F1}}{Q_1}\\
        \frac{w^i_{D1}}{Q_5}
    \end{bmatrix}
    = \begin{bmatrix}
        a Q_5 & 1\\
        b Q_1 & 1
    \end{bmatrix}
    \begin{bmatrix}
        \frac{w^i}{N}\\
        \hat w^i
    \end{bmatrix}, \quad
    \begin{bmatrix}
        \frac{{\tilde p}_j}{Q_1}\\
        \frac{w_j^{D3}}{Q_5}
    \end{bmatrix} = 
    \begin{bmatrix}
        Q_5 & -b \\
        -Q_1 & a
    \end{bmatrix} 
    \begin{bmatrix}
        \frac{p_j}{N}\\
        \hat p_j
    \end{bmatrix}
    . 
\end{equation}
The matrices appearing here also have determinant $1$. Therefore, in terms of the $(Q_1,Q_5)$ NS-NS theory, the BPS sub-lattice $w,p = 0 \text{ mod } N$ is exactly
\begin{equation}
\begin{split}
    w_\text{F1}^i &= {\tilde p}_j = 0 \text{ mod } Q_1,\\
    w_\text{D1}^i &= w^\text{D3}_j = 0 \text{ mod }Q_5.
\end{split}
\end{equation}
Namely, the charges are all integer multiples of the corresponding Chern-Simons levels. States carrying other charges, including now also singly-wrapped D-branes, are not BPS.

We now discuss the bulk interpretation of this BPS lattice. Already in the undeformed case, the interpretation of the $\hat T^4$ charged states seems to have a paradoxical nature. On the one hand, they carry R-R charge ($\hat p$, $\hat w$), with the interpretation of D-branes. On the other hand, these states exactly decouple from the entire closed string spectrum (and for example don't emit gravitons). This seems to contradict the basic property of D-branes as worldsheet boundary conditions. For $Q_5>1$ we found a generalization of that question. The BPS states don't just saturate the bound. As they live entirely in the Sugawara CFT, they also decouple exactly (in correlation functions) from operators in the coset CFT. These modes appear to be completely topological, and yet they carry ($Q_5$-quantized) R-R charges like D-branes.\footnote{For $Q_5>1$ there are also operators with $O(1)$ R-R charge that are not BPS, and we expect them to take the form of D-branes in the bulk. For $Q_5=1$ all the operators with R-R charge are BPS, and in this theory we expect no D-brane particle excitations (see also \cite{Gaberdiel:2021kkp, Knighton:2024noc}).} Similarly, we find states carrying the same charge as $Q_1$ winding fundamental strings, that decouple from all operators in the coset CFT.

The solution to the paradox is that these states are pure bulk gauge modes, not D-branes, that nevertheless carry R-R (and/or NS-NS) charges. One way to think of these modes is as boundary modes of the low-energy Chern-Simons theories, as in our discussion of the $Q_5=1$ case above. Alternatively, the existence of such charged modes in string theory in the presence of non-zero Kalb-Ramond $H_3$ flux was described in \cite{Maldacena:2001ss}. Consider the $AdS_3\times S^3 \times T^4$ type IIA background with $H_3 = Q_5 \omega_3$, with $\omega_3$ the unit volume-form on the $S^3$ (the $AdS_3$ component of the flux is not important for our analysis).\footnote{We use the conventions of \cite{Maldacena:2001ss} which are slightly different than \cite{Polchinski:1998rr}, but more natural for these purposes.} 
Denote the pure-gauge mode of the $C_3$ field on $S^3$ by $\psi$,
\begin{equation}
    C_3 = \psi\  \omega_3 + c_3,
\end{equation}
where $c_3$ includes the rest of the directions. In type IIA, the gauge transformation of $\delta C_1 = d \Lambda$ is followed by $\delta C_3 =\Lambda H_3$ \cite{Polchinski:1998rr}, with 0-form $\Lambda$. In our case,
\begin{equation}
    \delta \psi = Q_5 \Lambda.
\end{equation}
Because $\psi$ is the $C_3$ holonomy over $S^3$, $\psi\sim \psi +2\pi$. The holonomy $\exp(i \psi)$ is thus an operator with charge $Q_5$ under the R-R gauge field $C_1$. In terms of type IIB charges (upon  T-dualizing one of the $T^4$ directions), those are pure $C_4$ (wrapping $S^3\times T^1$) gauge modes with $w_\text{D1}=Q_5$ R-R charge. 
Applying T-duality on the four directions of the torus also reveals pure $\tilde C_6$ modes (wrapping $S^3\times T^3$) with $Q_5$-quantized $w^\text{D3}$ charge. Applying $ST_{1234}S$ gives the topological modes with $Q_1$-quantized momentum and winding modes.
In the language of \cite{Maldacena:2001ss}, the full string theory includes a $U(1)^{16}$ gauge theory with $\mathbb{Z}^{16}$ charges, but the coset has only $\mathbb{Z}_{Q_1}^8 \oplus \mathbb{Z}_{Q_5}^8 \equiv \mathbb{Z}_{Q_1 Q_5}^{8}$ charges.

\section*{Acknowledgments}

We would like to thank Vijay Balasubramanian, Lorenz Eberhardt, Rajesh Gopakumar, David Kutasov, Juan Maldacena, Emil Martinec, Sameer Murthy, Mukund Rangamani, and Edward Witten for useful discussions. We especially thank Matthias Gaberdiel and Rohit Kalloor for many useful discussions and for collaboration in the early stages of the project.
This work was supported in part by an Israel Science Foundation (ISF) grant number 2159/22, by Simons Foundation grant 994296 (Simons
Collaboration on Confinement and QCD Strings), by the Minerva foundation with funding from the Federal German Ministry for Education and Research, by the German Research Foundation through a German-Israeli Project Cooperation (DIP) grant ``Holography and the Swampland'', and by a research grant from Martin Eisenstein. OA is the Samuel Sebba Professorial Chair of Pure and Applied Physics.

\appendix

\section{Toroidal sigma models and \titlemath{$J\bar J$} deformations}\label{app:orbifold}
In this appendix we write down basic facts about $1+1$ dimensional sigma models on tori and their $J\bar J$ deformations.
\subsection{Compact bosons}
We will follow Polchinski's notations \cite{Polchinski:1998rq} with
\begin{equation}
\begin{split}
    z &= \sigma^1 + i \sigma^2, \quad \bar z = \sigma^1 - i \sigma^2,\\
    \partial &= \frac{1}{2}(\partial_1 - i \partial_2), \quad \bar\partial = \frac{1}{2}(\partial_1 + i \partial_2),
\end{split}
\end{equation}
and $d^2 z = 2 d^2 \sigma$. We take $d$ compact scalar fields $X^i$ ($i=1,\cdots,d$) with
\begin{equation}
    X^i \sim X^i + 2\pi.
\end{equation}
In these conventions, we take the (Euclidean) action
\begin{equation}
\begin{split}
    S &= \frac{1}{4\pi} \int d^2 \sigma \left(G_{ij} g^{\mu\nu} + i B_{ij} \epsilon^{\mu\nu}\right) \partial_\mu X^i \partial_\nu X^j \\
    &= \frac{1}{2\pi} \int d^2 z \left(G_{ij} \partial X^i \bar\partial X^j + B_{ij} \partial X^i \bar\partial X^j\right),
\end{split}
\end{equation}
with some constant metric and B-field $E_{ij} = G_{ij} + B_{ij}$. 
The momentum and winding Noether currents are
\begin{equation}
\begin{split}
    J^{\mu}_i &= \frac{1}{2\pi} \left(G_{ij} g^{\mu\nu} + i B_{ij} \epsilon^{\mu\nu}\right)\partial_\nu X^j,\\
    \tilde J_\mu^i &= \epsilon_{\mu\nu} G^{ij} J_{j}^\nu = 
    \frac{1}{2\pi} \left(\delta_j^i \epsilon_{\mu\nu} - i G^{ik} B_{kj} g_{\mu\nu} \right)\partial^\nu X^j.
\end{split}    
\end{equation}
We can write these in a holomorphic/anti-holomorphic basis
\begin{equation}\label{eq:chiral_curr}
\begin{split}   
    J_{i} &= \frac{1}{2}(J_{i, z} - i \tilde J_{i, z})
    = \frac{1}{2\pi} \left(
    G_{ij} - B_{ij} \right) \partial X^j,
    \\
    \bar J_{i} &= \frac{1}{2}(J_{i,\bar z} + i \tilde J_{i,\bar z}) = 
    \frac{1}{2\pi} \left(
    G_{ij} + B_{ij} \right) \bar \partial X^j,
\end{split}
\end{equation}
which satisfy $\bar \partial J_{i} = \partial \bar J_{i} = 0$. 

To consider the Hilbert space on a circle, we use the transformation $z=\exp(-i w)$, with
\begin{equation}
    w = \theta + i \tau \sim w + 2\pi.
\end{equation}
To consider the Lorentzian version we further take $\tau=it$. The different particle sectors are labeled by the winding $w^j \in \mathbb{Z}$
\begin{equation}
    X^j(w + 2 \pi) = X^j(w) + 2\pi w^j,
\end{equation}
and the center of mass momentum $p_i \in \mathbb{Z}$
\begin{equation}
    p_i = \int d\theta J^{t}_{i} = G_{ij} v^j + B_{ij} w^j,
\end{equation}
with the target space velocity $v^i$ , with the quantization
\begin{equation}
    v_i = p_i - B_{ij} w^j.
\end{equation}
The Hilbert space is labeled by the integers $p_i, w^j$ together with the usual harmonic oscillators $\alpha_n$.
The Hamiltonian (ignoring the Casimir energy) is given by
\begin{equation}\label{eq:full_H}
    H = \frac{1}{4}\left(v_R^2 + v_L^2\right) + N + \bar N,
\end{equation}
with
\begin{equation}
    v_{i,L/R} = v_i \pm G_{ij} w^j.
\end{equation}
The Sugawara energy is given solely by the first term in \eqref{eq:full_H},
which gives a lower bound on $L_0,\bar L_0$,
\begin{equation}\label{eq:energy_bound_sigma}
\begin{split}
    E^{L/R}_\text{Sugawara} = \frac{1}{4} v_{L/R}^2
    &= \frac{1}{4}v^2 + \frac{1}{4}w^2 \pm \frac{1}{2} v \cdot w\\
    &=\frac{1}{4}
    \begin{bmatrix}
        w^i & p_j
    \end{bmatrix}
    \begin{bmatrix}
        G + B^T G^{-1} B & \pm1 +B G^{-1}\\
        \pm 1-G^{-1}B & G^{-1}
    \end{bmatrix}
    \begin{bmatrix}
        w^i\\
        p_j 
    \end{bmatrix},
\end{split}
\end{equation}
where we omitted the $G$ contractions in the first line. The last formula is explicitly invariant under T-duality, which acts as
\begin{equation}
    p_i \leftrightarrow w^i, \quad E \leftrightarrow E^{-1}.
\end{equation}


Back to the Euclidean plane $z,\bar z$. For a given $G,B$ we would like to know the metric on the currents. 
The two-point function 
\begin{equation}
\begin{split}
    \langle \partial X^i(z) \partial X^j(0) \rangle &=  \frac{G^{ij}}{2z^2},\\
    \langle \bar\partial X^i(\bar z) \bar\partial X^j(0) \rangle &= \frac{G^{ij} }{2\bar z^2},
\end{split}
\end{equation}
gives the current two-point function for the chiral components of the momentum current \eqref{eq:chiral_curr}
\begin{equation}
\begin{split}
    \langle J_i(z) J_j(0) \rangle &= \frac{1}{8\pi^2} \frac{G_{ij} + B_{ik} G^{k l} B_{lj}}{z^2},\\
    \langle \bar J_i(\bar z) \bar J_j(0) \rangle &= 
    \frac{1}{8\pi^2} \frac{G_{ij} + B_{ik} G^{k l} B_{lj}}{\bar z^2}.
\end{split}
\end{equation}
This exactly the $G+B^T G^{-1}B$ term that we have in \eqref{eq:energy_bound_sigma}.
The winding Noether current is given by $\hat J^{\mu,i}=\frac{1}{2\pi}\epsilon^{\mu\nu}\partial_\mu X^i$. Its holomorphic and anti-holomorphic components are \footnote{These $i$'s are an outcome of the holomorphic coordinates in Euclidean space with $\epsilon_{z,\bar z} = \frac{i}{2}$. After continuing to the circle $w,\bar w = \theta \mp t$ we have $\epsilon_{w,\bar w} = \frac{1}{2}$, which gives the familiar formulas.}
\begin{equation}
    J^i = \frac{i}{2\pi}\partial X^i, \quad  \bar J^j = -\frac{i}{2\pi} \bar \partial X^j,
\end{equation}
which are also given by raising the target space index of \eqref{eq:chiral_curr} using $E$.
The two-point function is
\begin{equation}
\begin{split}
    \langle J^i(z) J^j(0) \rangle &= -\frac{1}{8\pi^2} \frac{G^{ij}}{z^2},\\
    \langle {\bar J}^i(\bar z) {\bar J}^j(0) \rangle &= 
    -\frac{1}{8\pi^2} \frac{G^{ij}}{\bar z^2},
\end{split}
\end{equation}
with the mixed terms
\begin{equation}
\begin{split}
    \langle J_i(z) J^j(0) \rangle &= \frac{i}{8\pi^2} \frac{\delta_i^j-B_{ik}G^{kj}}{z^2},\\
    \langle \bar J_i(\bar z) {\bar J}^j(0) \rangle &= 
    -\frac{i}{8\pi^2} \frac{\delta_i^j+B_{ik}G^{kj}}{\bar z^2}.
\end{split}
\end{equation}
We can directly see that the matrix \eqref{eq:energy_bound_sigma} is also the matrix of the current-current two-point functions, as required in the Sugawara construction.

The $J \bar J$ deformation \cite{Giveon:1994fu} (see also the recent \cite{Borsato:2023dis}) is given by the general deformation
\begin{equation} \label{eq:JJ_def_sigma}
    \partial_\lambda S  = 2\pi \int d^2 z M^{ij} J_i \bar J_j
    =\frac{1}{2\pi}\int d^2z E_{ki} M^{ij} E_{jl} \partial X^k \bar \partial X^l,
\end{equation}
for some matrix $M^{ij}$ and currents $J_i, \bar J_j$
depending
on $\lambda$. In other words, the sigma-model moduli exactly changes with
\begin{equation}\label{eq:change_G}
    \partial_\lambda E_{ij} = E_{ik} M^{kl} E_{lj}.
\end{equation}
These deformations cover the entire space of $E_{ij}$ deformations. They can be labeled by the $O(d,d)/\\(O(d)\times O(d))$ moduli.

\subsection{Going to the next level}

We will now specialize to the (bosonic version of) theory \eqref{eq:freeorb}. Namely, we have $4N$ fields $X^{I,i}$ ($I=1,\cdots,N$ and $i=1,\cdots,4$) which will constitute the $T^{4N}/S_N$ orbifold, and $4$ more fields $\hat X^i$. 
We assume the moduli on the $X^{I,i}$ are independent of $I$ and consist of ``single-trace" moduli. The action is given by
\begin{equation}\label{eq:full_action}
\begin{split}
    S &= \frac{1}{2\pi} \int d^2 z \left(E_{ij} \partial X^{I,i} \bar\partial X^{I,j} 
    + \hat E_{ij} \partial \hat X^i \bar\partial \hat X^j\right).
\end{split}
\end{equation}

The single-trace momentum currents are
\begin{equation}\label{eq:currents}
\begin{split}
    J_i = \frac{1}{2\pi} \sum_{I=1}^N E_{ji} \partial X^{I,j}, &\quad
    \bar J_i = \frac{1}{2\pi} \sum_{I=1}^N E_{ij} \bar \partial X^{I,j},\\
    \hat J_i = \frac{1}{2\pi}\hat E_{ji} \partial \hat X^{j}, &\quad
    \hat {\bar J}_i = \frac{1}{2\pi} \hat E_{ij} \bar \partial \hat X^{j}.
\end{split}
\end{equation}
In the normalization we use, the $w$-cycle twisted sectors (including the untwisted sector) have integer charges $Q_i, Q^j$. We denote the integer charges of these currents by $w^i$, $p_j$, $\hat w^k$ and $\hat p_l$, respectively.

Setting $E=G+B$, $\hat E = \hat G + \hat B$, the two-point function for the currents is
\begin{equation}\label{eq:JJ_orb}
\begin{split}
    \langle J_i(z) J_j(0) \rangle &= \frac{N}{8\pi^2 }\frac{G_{ij} + B_{ik} G^{kl}B_{lj}}{z^2},\\
    \langle \hat J_i(z) \hat J_j(0) \rangle &= \frac{1}{8\pi^2}\frac{\hat G_{ij} + \hat B_{ik} \hat G^{kl} \hat B_{lj}}{z^2}.
\end{split}
\end{equation}
Inverting the matrix, the Sugawara energy bound is given by
\begin{equation}\label{eq:energy_bound_orb}
\begin{split}
    E^{L/R}_\text{Sugawara} &=\frac{1}{4}
    \begin{bmatrix}
        w & p & \hat w & \hat p
    \end{bmatrix}
    \begin{bmatrix}
        \frac{G+B^T G^{-1} B}{N} & \frac{\pm 1 + B G^{-1}}{N} & 0 & 0\\
        \frac{\pm 1 - G^{-1} B}{N} & \frac{1}{N} G^{-1} & 0 & 0\\
        0 & 0 & \hat G + \hat B^T \hat G^{-1} \hat B & \pm 1 + \hat B \hat G^{-1}\\
        0 & 0 & \pm 1 -\hat G^{-1} \hat B & \hat G^{-1}
    \end{bmatrix}
    \begin{bmatrix}
        w\\
        p \\
        \hat w\\
        \hat p
    \end{bmatrix}.
\end{split}
\end{equation}
Notice that this is exactly two blocks of \eqref{eq:energy_bound_sigma}, only with the $p, w$ block multiplied by $1/N$ (as the two-point functions of $J$ are proportional to $N$).

A general $J\bar J$ deformation in this theory can be written as
\begin{equation} \label{eq:JJ_def_orb}
    \partial_\lambda S  = 2\pi
    \int d^2 z \left(
    M_{ij}^{11} J^i \bar J^j + 
    M_{ij}^{12} J^i \hat {\bar J}^j + 
    M_{ij}^{21} \hat J^i \bar J^j + 
    M_{ij}^{22} \hat J^i \hat {\bar J}^j
    \right)
    ,
\end{equation}
for some $8\times 8$ matrix $M_{ij}^{\alpha\beta}$ ($\alpha,\beta = 1,2$, $i,j = 1,\cdots,4$). In \eqref{eq:JJ_def_orb} we raise indices of $J,\bar J$ by $E$, and those of $\hat J, \hat {\bar J}$ by $\hat E$.
The $M^{22}$ matrix simply deforms the $\hat T^4$ moduli just like \eqref{eq:change_G},
\begin{equation}\label{eq:change_hatG}
    \partial_\lambda \hat E_{ij} = M_{ij}^{22}.
\end{equation}
$M^{11}, M^{12}$ and $M^{21}$ do not have that interpretation anymore, but together with $M^{22}$ they label an $SO(8,8)/(O(8)\times O(8))$ manifold.

To make contact with the $\chi$-deformation of \eqref{eq:bps_chi}, we set $B=\hat B=0$ and consider the $J \bar J$ deformation
\begin{equation}
\begin{split}
    M^{11}_{ij} &= 0,\\
    M^{12}_{ij} &= M^{21}_{ij} = - \frac{\chi}{N} G_{ij},\\
    M^{22}_{ij} &= \frac{V}{N} \chi^2 \hat G_{ij} = 
    \frac{\chi^2}{N} G_{ij},
\end{split}
\end{equation}
where we took $\hat G = \frac{1}{V} G$ as in section \ref{sec:decoupled_sec}. In terms of the (un-gauged) $4N+4$ fields, we deformed the $(4N+4)\times(4N+4)$ metric to
\begin{equation}
\begin{split}
    G_{4N+4} = 
    \mleft[
    \begin{array}{c|c}
      \delta_{IJ} & -  \chi \frac{1}{N}\\
      \hline
      -  \chi \frac{1}{N} & V^{-1} + \frac{\chi^2}{N}
    \end{array}
    \mright] \otimes G,
\end{split}
\end{equation}
with the inverse 
\begin{equation}
\begin{split}
    G_{4N+4}^{-1} = 
    \mleft[
    \begin{array}{c|c}
      \delta_{IJ} + \frac{V \chi^2}{N^2} & \chi \frac{V}{N} \\
      \hline
       \chi \frac{V}{N} & V
    \end{array}
    \mright] \otimes G^{-1}.
\end{split}
\end{equation}
Projecting to the single-trace currents \eqref{eq:currents} gives the Sugawara energy
\begin{equation}\label{eq:energy_bound_chi}
\begin{split}
    E^{L/R}_\text{Sugawara} &=\frac{1}{4}
    \begin{bmatrix}
        w & \hat w & p & \hat p
    \end{bmatrix}
    \begin{bmatrix}
        \frac{1}{N} G & - \chi \frac{1}{N} G & \pm \frac{1}{N} & 0\\
        - \chi \frac{1}{N} G & \frac{1+ \frac{V}{N} \chi^2}{V} G & 0 & \pm 1\\
        \pm \frac{1}{N} & 0 & \frac{1+ \frac{V}{N} \chi^2}{N} G^{-1} & \chi \frac{V}{N} G^{-1}\\
        0 & \pm 1 & \chi \frac{V}{N} G^{-1} & V G^{-1}
    \end{bmatrix}
    \begin{bmatrix}
        w \\
        \hat w\\
        p \\
        \hat p
    \end{bmatrix}.
\end{split}
\end{equation}
Re-writing in terms of contractions with the unit metric $V^{-1/2} G$ (and changing the order of the vector components) gives exactly \eqref{eq:bps_chi}.

\bibliographystyle{JHEP}
\bibliography{refs}

\providecommand{\href}[2]{#2}\begingroup\raggedright\begin{thebibliography}{10}

\bibitem{Maldacena:1997re}
J.M.~Maldacena, \emph{{The Large N limit of superconformal field theories and
  supergravity}}, \href{https://doi.org/10.4310/ATMP.1998.v2.n2.a1}{\emph{Adv.
  Theor. Math. Phys.} {\bfseries 2} (1998) 231}
  [\href{https://arxiv.org/abs/hep-th/9711200}{{\ttfamily hep-th/9711200}}].

\bibitem{Aharony:1999ti}
O.~Aharony, S.S.~Gubser, J.M.~Maldacena, H.~Ooguri and Y.~Oz, \emph{{Large N
  field theories, string theory and gravity}},
  \href{https://doi.org/10.1016/S0370-1573(99)00083-6}{\emph{Phys. Rept.}
  {\bfseries 323} (2000) 183}
  [\href{https://arxiv.org/abs/hep-th/9905111}{{\ttfamily hep-th/9905111}}].

\bibitem{Maldacena:1998bw}
J.M.~Maldacena and A.~Strominger, \emph{{AdS(3) black holes and a stringy
  exclusion principle}},
  \href{https://doi.org/10.1088/1126-6708/1998/12/005}{\emph{JHEP} {\bfseries
  12} (1998) 005} [\href{https://arxiv.org/abs/hep-th/9804085}{{\ttfamily
  hep-th/9804085}}].

\bibitem{Evans:1998qu}
J.M.~Evans, M.R.~Gaberdiel and M.J.~Perry, \emph{{The no ghost theorem for
  AdS(3) and the stringy exclusion principle}},
  \href{https://doi.org/10.1016/S0550-3213(98)00561-6}{\emph{Nucl. Phys. B}
  {\bfseries 535} (1998) 152}
  [\href{https://arxiv.org/abs/hep-th/9806024}{{\ttfamily hep-th/9806024}}].

\bibitem{Giveon:1998ns}
A.~Giveon, D.~Kutasov and N.~Seiberg, \emph{{Comments on string theory on
  AdS(3)}}, \href{https://doi.org/10.4310/ATMP.1998.v2.n4.a3}{\emph{Adv. Theor.
  Math. Phys.} {\bfseries 2} (1998) 733}
  [\href{https://arxiv.org/abs/hep-th/9806194}{{\ttfamily hep-th/9806194}}].

\bibitem{deBoer:1998gyt}
J.~de~Boer, H.~Ooguri, H.~Robins and J.~Tannenhauser, \emph{{String theory on
  AdS(3)}}, \href{https://doi.org/10.1088/1126-6708/1998/12/026}{\emph{JHEP}
  {\bfseries 12} (1998) 026}
  [\href{https://arxiv.org/abs/hep-th/9812046}{{\ttfamily hep-th/9812046}}].

\bibitem{Kutasov:1999xu}
D.~Kutasov and N.~Seiberg, \emph{{More comments on string theory on AdS(3)}},
  \href{https://doi.org/10.1088/1126-6708/1999/04/008}{\emph{JHEP} {\bfseries
  04} (1999) 008} [\href{https://arxiv.org/abs/hep-th/9903219}{{\ttfamily
  hep-th/9903219}}].

\bibitem{Giveon:2001up}
A.~Giveon and D.~Kutasov, \emph{{Notes on AdS(3)}},
  \href{https://doi.org/10.1016/S0550-3213(01)00573-9}{\emph{Nucl. Phys. B}
  {\bfseries 621} (2002) 303}
  [\href{https://arxiv.org/abs/hep-th/0106004}{{\ttfamily hep-th/0106004}}].

\bibitem{Cho:2018nfn}
M.~Cho, S.~Collier and X.~Yin, \emph{{Strings in Ramond-Ramond Backgrounds from
  the Neveu-Schwarz-Ramond Formalism}},
  \href{https://doi.org/10.1007/JHEP12(2020)123}{\emph{JHEP} {\bfseries 12}
  (2020) 123} [\href{https://arxiv.org/abs/1811.00032}{{\ttfamily
  1811.00032}}].

\bibitem{Giribet:2018ada}
G.~Giribet, C.~Hull, M.~Kleban, M.~Porrati and E.~Rabinovici,
  \emph{{Superstrings on AdS$_{3}$ at $k = 1$}},
  \href{https://doi.org/10.1007/JHEP08(2018)204}{\emph{JHEP} {\bfseries 08}
  (2018) 204} [\href{https://arxiv.org/abs/1803.04420}{{\ttfamily
  1803.04420}}].

\bibitem{Berkovits:1999im}
N.~Berkovits, C.~Vafa and E.~Witten, \emph{{Conformal field theory of AdS
  background with Ramond-Ramond flux}},
  \href{https://doi.org/10.1088/1126-6708/1999/03/018}{\emph{JHEP} {\bfseries
  03} (1999) 018} [\href{https://arxiv.org/abs/hep-th/9902098}{{\ttfamily
  hep-th/9902098}}].

\bibitem{Gaberdiel:2011vf}
M.R.~Gaberdiel and S.~Gerigk, \emph{{The massless string spectrum on AdS$_3$ x
  S$^3$ from the supergroup}},
  \href{https://doi.org/10.1007/JHEP10(2011)045}{\emph{JHEP} {\bfseries 10}
  (2011) 045} [\href{https://arxiv.org/abs/1107.2660}{{\ttfamily 1107.2660}}].

\bibitem{Eberhardt:2018ouy}
L.~Eberhardt, M.R.~Gaberdiel and R.~Gopakumar, \emph{{The Worldsheet Dual of
  the Symmetric Product CFT}},
  \href{https://doi.org/10.1007/JHEP04(2019)103}{\emph{JHEP} {\bfseries 04}
  (2019) 103} [\href{https://arxiv.org/abs/1812.01007}{{\ttfamily
  1812.01007}}].

\bibitem{Gaberdiel:2021njm}
M.R.~Gaberdiel and K.~Naderi, \emph{{The physical states of the Hybrid
  Formalism}}, \href{https://doi.org/10.1007/JHEP10(2021)168}{\emph{JHEP}
  {\bfseries 10} (2021) 168}
  [\href{https://arxiv.org/abs/2106.06476}{{\ttfamily 2106.06476}}].

\bibitem{Larsen:1999uk}
F.~Larsen and E.J.~Martinec, \emph{{U(1) charges and moduli in the D1 - D5
  system}}, \href{https://doi.org/10.1088/1126-6708/1999/06/019}{\emph{JHEP}
  {\bfseries 06} (1999) 019}
  [\href{https://arxiv.org/abs/hep-th/9905064}{{\ttfamily hep-th/9905064}}].

\bibitem{Dabholkar:2007ey}
A.~Dabholkar and A.~Pakman, \emph{{Exact chiral ring of AdS(3) / CFT(2)}},
  \href{https://doi.org/10.4310/ATMP.2009.v13.n2.a2}{\emph{Adv. Theor. Math.
  Phys.} {\bfseries 13} (2009) 409}
  [\href{https://arxiv.org/abs/hep-th/0703022}{{\ttfamily hep-th/0703022}}].

\bibitem{deBoer:2008ss}
J.~de~Boer, J.~Manschot, K.~Papadodimas and E.~Verlinde, \emph{{The Chiral ring
  of AdS(3)/CFT(2) and the attractor mechanism}},
  \href{https://doi.org/10.1088/1126-6708/2009/03/030}{\emph{JHEP} {\bfseries
  03} (2009) 030} [\href{https://arxiv.org/abs/0809.0507}{{\ttfamily
  0809.0507}}].

\bibitem{Lunin:2001pw}
O.~Lunin and S.D.~Mathur, \emph{{Three point functions for M(N) / S(N)
  orbifolds with N=4 supersymmetry}},
  \href{https://doi.org/10.1007/s002200200638}{\emph{Commun. Math. Phys.}
  {\bfseries 227} (2002) 385}
  [\href{https://arxiv.org/abs/hep-th/0103169}{{\ttfamily hep-th/0103169}}].

\bibitem{Pakman:2007hn}
A.~Pakman and A.~Sever, \emph{{Exact N=4 correlators of AdS(3)/CFT(2)}},
  \href{https://doi.org/10.1016/j.physletb.2007.06.041}{\emph{Phys. Lett. B}
  {\bfseries 652} (2007) 60} [\href{https://arxiv.org/abs/0704.3040}{{\ttfamily
  0704.3040}}].

\bibitem{Pakman:2009ab}
A.~Pakman, L.~Rastelli and S.S.~Razamat, \emph{{Extremal Correlators and
  Hurwitz Numbers in Symmetric Product Orbifolds}},
  \href{https://doi.org/10.1103/PhysRevD.80.086009}{\emph{Phys. Rev. D}
  {\bfseries 80} (2009) 086009}
  [\href{https://arxiv.org/abs/0905.3451}{{\ttfamily 0905.3451}}].

\bibitem{Baggio:2012rr}
M.~Baggio, J.~de~Boer and K.~Papadodimas, \emph{{A non-renormalization theorem
  for chiral primary 3-point functions}},
  \href{https://doi.org/10.1007/JHEP07(2012)137}{\emph{JHEP} {\bfseries 07}
  (2012) 137} [\href{https://arxiv.org/abs/1203.1036}{{\ttfamily 1203.1036}}].

\bibitem{Gaberdiel:2022oeu}
M.R.~Gaberdiel and B.~Nairz, \emph{{BPS correlators for AdS$_{3}$/CFT$_{2}$}},
  \href{https://doi.org/10.1007/JHEP09(2022)244}{\emph{JHEP} {\bfseries 09}
  (2022) 244} [\href{https://arxiv.org/abs/2207.03956}{{\ttfamily
  2207.03956}}].

\bibitem{Martinec:2022okx}
E.J.~Martinec, S.~Massai and D.~Turton, \emph{{On the BPS Sector in AdS3/CFT2
  Holography}}, \href{https://doi.org/10.1002/prop.202300015}{\emph{Fortsch.
  Phys.} {\bfseries 71} (2023) 2300015}
  [\href{https://arxiv.org/abs/2211.12476}{{\ttfamily 2211.12476}}].

\bibitem{Iguri:2023khc}
S.~Iguri, N.~Kovensky and J.H.~Toro, \emph{{Spectral flow and the exact
  AdS$_{3}$/CFT$_{2}$ chiral ring}},
  \href{https://doi.org/10.1007/JHEP08(2023)034}{\emph{JHEP} {\bfseries 08}
  (2023) 034} [\href{https://arxiv.org/abs/2304.08361}{{\ttfamily
  2304.08361}}].

\bibitem{Giribet:2001ft}
G.~Giribet and C.A.~Nunez, \emph{{Correlators in AdS(3) string theory}},
  \href{https://doi.org/10.1088/1126-6708/2001/06/010}{\emph{JHEP} {\bfseries
  06} (2001) 010} [\href{https://arxiv.org/abs/hep-th/0105200}{{\ttfamily
  hep-th/0105200}}].

\bibitem{Maldacena:2001km}
J.M.~Maldacena and H.~Ooguri, \emph{{Strings in AdS(3) and the SL(2,R) WZW
  model. Part 3. Correlation functions}},
  \href{https://doi.org/10.1103/PhysRevD.65.106006}{\emph{Phys. Rev. D}
  {\bfseries 65} (2002) 106006}
  [\href{https://arxiv.org/abs/hep-th/0111180}{{\ttfamily hep-th/0111180}}].

\bibitem{Troost:2002wk}
J.~Troost, \emph{{Winding strings and AdS(3) black holes}},
  \href{https://doi.org/10.1088/1126-6708/2002/09/041}{\emph{JHEP} {\bfseries
  09} (2002) 041} [\href{https://arxiv.org/abs/hep-th/0206118}{{\ttfamily
  hep-th/0206118}}].

\bibitem{Gaberdiel:2007vu}
M.R.~Gaberdiel and I.~Kirsch, \emph{{Worldsheet correlators in AdS(3)/CFT(2)}},
  \href{https://doi.org/10.1088/1126-6708/2007/04/050}{\emph{JHEP} {\bfseries
  04} (2007) 050} [\href{https://arxiv.org/abs/hep-th/0703001}{{\ttfamily
  hep-th/0703001}}].

\bibitem{Giribet:2007wp}
G.~Giribet, A.~Pakman and L.~Rastelli, \emph{{Spectral Flow in AdS(3)/CFT(2)}},
  \href{https://doi.org/10.1088/1126-6708/2008/06/013}{\emph{JHEP} {\bfseries
  06} (2008) 013} [\href{https://arxiv.org/abs/0712.3046}{{\ttfamily
  0712.3046}}].

\bibitem{Taylor:2007hs}
M.~Taylor, \emph{{Matching of correlators in AdS(3) / CFT(2)}},
  \href{https://doi.org/10.1088/1126-6708/2008/06/010}{\emph{JHEP} {\bfseries
  06} (2008) 010} [\href{https://arxiv.org/abs/0709.1838}{{\ttfamily
  0709.1838}}].

\bibitem{Cardona:2009hk}
C.A.~Cardona and C.A.~Nunez, \emph{{Three-point functions in superstring theory
  on AdS(3) x S**3 x T**4}},
  \href{https://doi.org/10.1088/1126-6708/2009/06/009}{\emph{JHEP} {\bfseries
  06} (2009) 009} [\href{https://arxiv.org/abs/0903.2001}{{\ttfamily
  0903.2001}}].

\bibitem{Cardona:2010qf}
C.A.~Cardona and I.~Kirsch, \emph{{Worldsheet four-point functions in
  $AdS_3/CFT_2$}}, \href{https://doi.org/10.1007/JHEP01(2011)015}{\emph{JHEP}
  {\bfseries 01} (2011) 015} [\href{https://arxiv.org/abs/1007.2720}{{\ttfamily
  1007.2720}}].

\bibitem{Dei:2021xgh}
A.~Dei and L.~Eberhardt, \emph{{String correlators on AdS$_{3}$: three-point
  functions}}, \href{https://doi.org/10.1007/JHEP08(2021)025}{\emph{JHEP}
  {\bfseries 08} (2021) 025}
  [\href{https://arxiv.org/abs/2105.12130}{{\ttfamily 2105.12130}}].

\bibitem{Dei:2021yom}
A.~Dei and L.~Eberhardt, \emph{{String correlators on AdS$_{3}$: four-point
  functions}}, \href{https://doi.org/10.1007/JHEP09(2021)209}{\emph{JHEP}
  {\bfseries 09} (2021) 209}
  [\href{https://arxiv.org/abs/2107.01481}{{\ttfamily 2107.01481}}].

\bibitem{Dei:2022pkr}
A.~Dei and L.~Eberhardt, \emph{{String correlators on $\text{AdS}_3$: Analytic
  structure and dual CFT}},
  \href{https://doi.org/10.21468/SciPostPhys.13.3.053}{\emph{SciPost Phys.}
  {\bfseries 13} (2022) 053}
  [\href{https://arxiv.org/abs/2203.13264}{{\ttfamily 2203.13264}}].

\bibitem{Dei:2023ivl}
A.~Dei, B.~Knighton and K.~Naderi, \emph{{Solving AdS$_3$ string theory at
  minimal tension: tree-level correlators}},
  \href{https://arxiv.org/abs/2312.04622}{{\ttfamily 2312.04622}}.

\bibitem{Maldacena:2000hw}
J.M.~Maldacena and H.~Ooguri, \emph{{Strings in AdS(3) and SL(2,R) WZW model
  1.: The Spectrum}}, \href{https://doi.org/10.1063/1.1377273}{\emph{J. Math.
  Phys.} {\bfseries 42} (2001) 2929}
  [\href{https://arxiv.org/abs/hep-th/0001053}{{\ttfamily hep-th/0001053}}].

\bibitem{Kim:2015gak}
J.~Kim and M.~Porrati, \emph{{On the central charge of spacetime current
  algebras and correlators in string theory on AdS$_{3}$}},
  \href{https://doi.org/10.1007/JHEP05(2015)076}{\emph{JHEP} {\bfseries 05}
  (2015) 076} [\href{https://arxiv.org/abs/1503.07186}{{\ttfamily
  1503.07186}}].

\bibitem{Lunin:2000yv}
O.~Lunin and S.D.~Mathur, \emph{{Correlation functions for M**N / S(N)
  orbifolds}}, \href{https://doi.org/10.1007/s002200100431}{\emph{Commun. Math.
  Phys.} {\bfseries 219} (2001) 399}
  [\href{https://arxiv.org/abs/hep-th/0006196}{{\ttfamily hep-th/0006196}}].

\bibitem{Pakman:2009zz}
A.~Pakman, L.~Rastelli and S.S.~Razamat, \emph{{Diagrams for Symmetric Product
  Orbifolds}}, \href{https://doi.org/10.1088/1126-6708/2009/10/034}{\emph{JHEP}
  {\bfseries 10} (2009) 034} [\href{https://arxiv.org/abs/0905.3448}{{\ttfamily
  0905.3448}}].

\bibitem{Pakman:2009mi}
A.~Pakman, L.~Rastelli and S.S.~Razamat, \emph{{A Spin Chain for the Symmetric
  Product CFT(2)}}, \href{https://doi.org/10.1007/JHEP05(2010)099}{\emph{JHEP}
  {\bfseries 05} (2010) 099} [\href{https://arxiv.org/abs/0912.0959}{{\ttfamily
  0912.0959}}].

\bibitem{Eberhardt:2019ywk}
L.~Eberhardt, M.R.~Gaberdiel and R.~Gopakumar, \emph{{Deriving the
  AdS$_{3}$/CFT$_{2}$ correspondence}},
  \href{https://doi.org/10.1007/JHEP02(2020)136}{\emph{JHEP} {\bfseries 02}
  (2020) 136} [\href{https://arxiv.org/abs/1911.00378}{{\ttfamily
  1911.00378}}].

\bibitem{Dijkgraaf:1996xw}
R.~Dijkgraaf, G.W.~Moore, E.P.~Verlinde and H.L.~Verlinde, \emph{{Elliptic
  genera of symmetric products and second quantized strings}},
  \href{https://doi.org/10.1007/s002200050087}{\emph{Commun. Math. Phys.}
  {\bfseries 185} (1997) 197}
  [\href{https://arxiv.org/abs/hep-th/9608096}{{\ttfamily hep-th/9608096}}].

\bibitem{Bantay:2000eq}
P.~Bantay, \emph{{Symmetric products, permutation orbifolds and discrete
  torsion}}, \href{https://doi.org/10.1023/A:1024453119772}{\emph{Lett. Math.
  Phys.} {\bfseries 63} (2003) 209}
  [\href{https://arxiv.org/abs/hep-th/0004025}{{\ttfamily hep-th/0004025}}].

\bibitem{Eberhardt:2021jvj}
L.~Eberhardt, \emph{{Summing over Geometries in String Theory}},
  \href{https://doi.org/10.1007/JHEP05(2021)233}{\emph{JHEP} {\bfseries 05}
  (2021) 233} [\href{https://arxiv.org/abs/2102.12355}{{\ttfamily
  2102.12355}}].

\bibitem{Seiberg:1999xz}
N.~Seiberg and E.~Witten, \emph{{The D1 / D5 system and singular CFT}},
  \href{https://doi.org/10.1088/1126-6708/1999/04/017}{\emph{JHEP} {\bfseries
  04} (1999) 017} [\href{https://arxiv.org/abs/hep-th/9903224}{{\ttfamily
  hep-th/9903224}}].

\bibitem{Eberhardt:2019niq}
L.~Eberhardt and M.R.~Gaberdiel, \emph{{Strings on $\text{AdS}_3 \times
  \text{S}^3 \times \text{S}^3 \times \text{S}^1$}},
  \href{https://doi.org/10.1007/JHEP06(2019)035}{\emph{JHEP} {\bfseries 06}
  (2019) 035} [\href{https://arxiv.org/abs/1904.01585}{{\ttfamily
  1904.01585}}].

\bibitem{Kutasov:1998zh}
D.~Kutasov, F.~Larsen and R.G.~Leigh, \emph{{String theory in magnetic monopole
  backgrounds}},
  \href{https://doi.org/10.1016/S0550-3213(99)00144-3}{\emph{Nucl. Phys. B}
  {\bfseries 550} (1999) 183}
  [\href{https://arxiv.org/abs/hep-th/9812027}{{\ttfamily hep-th/9812027}}].

\bibitem{Martinec:2001cf}
E.J.~Martinec and W.~McElgin, \emph{{String theory on AdS orbifolds}},
  \href{https://doi.org/10.1088/1126-6708/2002/04/029}{\emph{JHEP} {\bfseries
  04} (2002) 029} [\href{https://arxiv.org/abs/hep-th/0106171}{{\ttfamily
  hep-th/0106171}}].

\bibitem{Martinec:2023zha}
E.J.~Martinec, \emph{{AdS$_{3}$ orbifolds, BTZ black holes, and holography}},
  \href{https://doi.org/10.1007/JHEP10(2023)016}{\emph{JHEP} {\bfseries 10}
  (2023) 016} [\href{https://arxiv.org/abs/2307.02559}{{\ttfamily
  2307.02559}}].

\bibitem{Gaberdiel:2023dxt}
M.R.~Gaberdiel, B.~Guo and S.D.~Mathur, \emph{{Tensionless strings on AdS$_{3}$
  orbifolds}}, \href{https://doi.org/10.1007/JHEP04(2024)057}{\emph{JHEP}
  {\bfseries 04} (2024) 057}
  [\href{https://arxiv.org/abs/2312.01348}{{\ttfamily 2312.01348}}].

\bibitem{Hohenegger:2008du}
S.~Hohenegger, C.A.~Keller and I.~Kirsch, \emph{{Heterotic AdS(3) / CFT(2)
  duality with (0,4) spacetime supersymmetry}},
  \href{https://doi.org/10.1016/j.nuclphysb.2008.06.020}{\emph{Nucl. Phys. B}
  {\bfseries 804} (2008) 193}
  [\href{https://arxiv.org/abs/0804.4066}{{\ttfamily 0804.4066}}].

\bibitem{Baykara:2022cwj}
Z.K.~Baykara, D.~Robbins and S.~Sethi, \emph{{Non-supersymmetric AdS from
  string theory}},
  \href{https://doi.org/10.21468/SciPostPhys.15.6.224}{\emph{SciPost Phys.}
  {\bfseries 15} (2023) 224}
  [\href{https://arxiv.org/abs/2212.02557}{{\ttfamily 2212.02557}}].

\bibitem{Fraiman:2023cpa}
B.~Fraiman, M.~Gra\~na, H.~Parra De~Freitas and S.~Sethi,
  \emph{{Non-Supersymmetric Heterotic Strings on a Circle}},
  \href{https://arxiv.org/abs/2307.13745}{{\ttfamily 2307.13745}}.

\bibitem{Aharony:1997th}
O.~Aharony, M.~Berkooz, S.~Kachru, N.~Seiberg and E.~Silverstein, \emph{{Matrix
  description of interacting theories in six-dimensions}},
  \href{https://doi.org/10.4310/ATMP.1997.v1.n1.a5}{\emph{Adv. Theor. Math.
  Phys.} {\bfseries 1} (1998) 148}
  [\href{https://arxiv.org/abs/hep-th/9707079}{{\ttfamily hep-th/9707079}}].

\bibitem{Witten:1997yu}
E.~Witten, \emph{{On the conformal field theory of the Higgs branch}},
  \href{https://doi.org/10.1088/1126-6708/1997/07/003}{\emph{JHEP} {\bfseries
  07} (1997) 003} [\href{https://arxiv.org/abs/hep-th/9707093}{{\ttfamily
  hep-th/9707093}}].

\bibitem{Aharony:1999dw}
O.~Aharony and M.~Berkooz, \emph{{IR dynamics of D = 2, N=(4,4) gauge theories
  and DLCQ of 'little string theories'}},
  \href{https://doi.org/10.1088/1126-6708/1999/10/030}{\emph{JHEP} {\bfseries
  10} (1999) 030} [\href{https://arxiv.org/abs/hep-th/9909101}{{\ttfamily
  hep-th/9909101}}].

\bibitem{Taubes:1983bk}
C.H.~Taubes, \emph{{Self-dual connections on 4-manifolds with indefinite
  intersection matrix}}, {\emph{J. Diff. Geom.} {\bfseries 19} (1984) 517}.

\bibitem{Gerigk:2012lqa}
S.~Gerigk, \emph{{Superstring theory on AdS$_3$ x S\textthreesuperior{} and the
  PSL$(2|2)$ WZW model}}, Ph.D. thesis, Zurich, ETH, 2012.
\newblock 10.3929/ethz-a-007595532.

\bibitem{Gaberdiel:2018rqv}
M.R.~Gaberdiel and R.~Gopakumar, \emph{{Tensionless string spectra on
  AdS$_{3}$}}, \href{https://doi.org/10.1007/JHEP05(2018)085}{\emph{JHEP}
  {\bfseries 05} (2018) 085}
  [\href{https://arxiv.org/abs/1803.04423}{{\ttfamily 1803.04423}}].

\bibitem{Dei:2020zui}
A.~Dei, M.R.~Gaberdiel, R.~Gopakumar and B.~Knighton, \emph{{Free field
  world-sheet correlators for ${\rm AdS}_3$}},
  \href{https://doi.org/10.1007/JHEP02(2021)081}{\emph{JHEP} {\bfseries 02}
  (2021) 081} [\href{https://arxiv.org/abs/2009.11306}{{\ttfamily
  2009.11306}}].

\bibitem{Eberhardt:2020akk}
L.~Eberhardt, \emph{{AdS$_{3}$/CFT$_{2}$ at higher genus}},
  \href{https://doi.org/10.1007/JHEP05(2020)150}{\emph{JHEP} {\bfseries 05}
  (2020) 150} [\href{https://arxiv.org/abs/2002.11729}{{\ttfamily
  2002.11729}}].

\bibitem{Knighton:2020kuh}
B.~Knighton, \emph{{Higher genus correlators for tensionless AdS$_{3}$
  strings}}, \href{https://doi.org/10.1007/JHEP04(2021)211}{\emph{JHEP}
  {\bfseries 04} (2021) 211}
  [\href{https://arxiv.org/abs/2012.01445}{{\ttfamily 2012.01445}}].

\bibitem{schoeneberg2012elliptic}
B.~Schoeneberg, J.~Smart and E.~Schwandt, \emph{Elliptic Modular Functions: An
  Introduction}, Grundlehren der mathematischen Wissenschaften, Springer Berlin
  Heidelberg (2012).

\bibitem{Argurio:2000tb}
R.~Argurio, A.~Giveon and A.~Shomer, \emph{{Superstrings on AdS(3) and
  symmetric products}},
  \href{https://doi.org/10.1088/1126-6708/2000/12/003}{\emph{JHEP} {\bfseries
  12} (2000) 003} [\href{https://arxiv.org/abs/hep-th/0009242}{{\ttfamily
  hep-th/0009242}}].

\bibitem{Eberhardt:2019qcl}
L.~Eberhardt and M.R.~Gaberdiel, \emph{{String theory on AdS$_3$ and the
  symmetric orbifold of Liouville theory}},
  \href{https://doi.org/10.1016/j.nuclphysb.2019.114774}{\emph{Nucl. Phys. B}
  {\bfseries 948} (2019) 114774}
  [\href{https://arxiv.org/abs/1903.00421}{{\ttfamily 1903.00421}}].

\bibitem{Dei:2019osr}
A.~Dei, L.~Eberhardt and M.R.~Gaberdiel, \emph{{Three-point functions in
  AdS$_{3}$/CFT$_{2}$ holography}},
  \href{https://doi.org/10.1007/JHEP12(2019)012}{\emph{JHEP} {\bfseries 12}
  (2019) 012} [\href{https://arxiv.org/abs/1907.13144}{{\ttfamily
  1907.13144}}].

\bibitem{Eberhardt:2021vsx}
L.~Eberhardt, \emph{{A perturbative CFT dual for pure NS\textendash{}NS
  AdS$_{3}$ strings}}, \href{https://doi.org/10.1088/1751-8121/ac47b2}{\emph{J.
  Phys. A} {\bfseries 55} (2022) 064001}
  [\href{https://arxiv.org/abs/2110.07535}{{\ttfamily 2110.07535}}].

\bibitem{Aspinwall:1995zi}
P.S.~Aspinwall, \emph{{Enhanced gauge symmetries and K3 surfaces}},
  \href{https://doi.org/10.1016/0370-2693(95)00957-M}{\emph{Phys. Lett. B}
  {\bfseries 357} (1995) 329}
  [\href{https://arxiv.org/abs/hep-th/9507012}{{\ttfamily hep-th/9507012}}].

\bibitem{Eberhardt:2020bgq}
L.~Eberhardt, \emph{{Partition functions of the tensionless string}},
  \href{https://doi.org/10.1007/JHEP03(2021)176}{\emph{JHEP} {\bfseries 03}
  (2021) 176} [\href{https://arxiv.org/abs/2008.07533}{{\ttfamily
  2008.07533}}].

\bibitem{Knighton:2024ybs}
B.~Knighton, \emph{{A note on background independence in $\text{AdS}_3$ string
  theory}},  \href{https://arxiv.org/abs/2404.19571}{{\ttfamily 2404.19571}}.

\bibitem{Knighton:2024noc}
B.~Knighton, V.~Sriprachyakul and J.~Vo\v{s}mera, \emph{{Topological defects
  and tensionless holography}},
  \href{https://arxiv.org/abs/2406.03467}{{\ttfamily 2406.03467}}.

\bibitem{Haehl:2014yla}
F.M.~Haehl and M.~Rangamani, \emph{{Permutation orbifolds and holography}},
  \href{https://doi.org/10.1007/JHEP03(2015)163}{\emph{JHEP} {\bfseries 03}
  (2015) 163} [\href{https://arxiv.org/abs/1412.2759}{{\ttfamily 1412.2759}}].

\bibitem{Belin:2015hwa}
A.~Belin, C.A.~Keller and A.~Maloney, \emph{{Permutation Orbifolds in the large
  N Limit}},  \href{https://arxiv.org/abs/1509.01256}{{\ttfamily 1509.01256}}.

\bibitem{Eberhardt:2023lwd}
L.~Eberhardt and S.~Pal, \emph{{Holographic Weyl anomaly in string theory}},
  \href{https://doi.org/10.21468/SciPostPhys.16.1.027}{\emph{SciPost Phys.}
  {\bfseries 16} (2024) 027}
  [\href{https://arxiv.org/abs/2307.03000}{{\ttfamily 2307.03000}}].

\bibitem{Gaberdiel:2020ycd}
M.R.~Gaberdiel, R.~Gopakumar, B.~Knighton and P.~Maity, \emph{{From symmetric
  product CFTs to AdS$_{3}$}},
  \href{https://doi.org/10.1007/JHEP05(2021)073}{\emph{JHEP} {\bfseries 05}
  (2021) 073} [\href{https://arxiv.org/abs/2011.10038}{{\ttfamily
  2011.10038}}].

\bibitem{Gukov:2004id}
S.~Gukov, E.~Martinec, G.W.~Moore and A.~Strominger, \emph{{Chern-Simons gauge
  theory and the AdS(3) / CFT(2) correspondence}},  in \emph{{From Fields to
  Strings: Circumnavigating Theoretical Physics: A Conference in Tribute to Ian
  Kogan}}, pp.~1606--1647, 3, 2004,
  \href{https://doi.org/10.1142/9789812775344_0036}{DOI}
  [\href{https://arxiv.org/abs/hep-th/0403225}{{\ttfamily hep-th/0403225}}].

\bibitem{Aharony:2023zit}
O.~Aharony, A.~Dymarsky and A.D.~Shapere, \emph{{Holographic description of
  Narain CFTs and their code-based ensembles}},
  \href{https://arxiv.org/abs/2310.06012}{{\ttfamily 2310.06012}}.

\bibitem{David:2002wn}
J.R.~David, G.~Mandal and S.R.~Wadia, \emph{{Microscopic formulation of black
  holes in string theory}},
  \href{https://doi.org/10.1016/S0370-1573(02)00271-5}{\emph{Phys. Rept.}
  {\bfseries 369} (2002) 549}
  [\href{https://arxiv.org/abs/hep-th/0203048}{{\ttfamily hep-th/0203048}}].

\bibitem{Benjamin:2021zkn}
N.~Benjamin, C.A.~Keller and I.G.~Zadeh, \emph{{Lifting 1/4-BPS states in
  AdS$_{3}$\texttimes{} S$^{3}$\texttimes{} T$^{4}$}},
  \href{https://doi.org/10.1007/JHEP10(2021)089}{\emph{JHEP} {\bfseries 10}
  (2021) 089} [\href{https://arxiv.org/abs/2107.00655}{{\ttfamily
  2107.00655}}].

\bibitem{Guo:2022ifr}
B.~Guo, M.R.R.~Hughes, S.D.~Mathur and M.~Mehta, \emph{{Universal lifting in
  the D1-D5 CFT}}, \href{https://doi.org/10.1007/JHEP10(2022)148}{\emph{JHEP}
  {\bfseries 10} (2022) 148}
  [\href{https://arxiv.org/abs/2208.07409}{{\ttfamily 2208.07409}}].

\bibitem{Apolo:2022fya}
L.~Apolo, A.~Belin, S.~Bintanja, A.~Castro and C.A.~Keller, \emph{{Deforming
  symmetric product orbifolds: a tale of moduli and higher spin currents}},
  \href{https://doi.org/10.1007/JHEP08(2022)159}{\emph{JHEP} {\bfseries 08}
  (2022) 159} [\href{https://arxiv.org/abs/2204.07590}{{\ttfamily
  2204.07590}}].

\bibitem{Guo:2022zpn}
B.~Guo and S.~Hampton, \emph{{Bootstrapping the effect of the twist operator in
  the D1D5 CFT}}, \href{https://doi.org/10.1007/JHEP03(2024)030}{\emph{JHEP}
  {\bfseries 03} (2024) 030}
  [\href{https://arxiv.org/abs/2210.07217}{{\ttfamily 2210.07217}}].

\bibitem{Fiset:2022erp}
M.-A.~Fiset, M.R.~Gaberdiel, K.~Naderi and V.~Sriprachyakul, \emph{{Perturbing
  the symmetric orbifold from the worldsheet}},
  \href{https://doi.org/10.1007/JHEP07(2023)093}{\emph{JHEP} {\bfseries 07}
  (2023) 093} [\href{https://arxiv.org/abs/2212.12342}{{\ttfamily
  2212.12342}}].

\bibitem{Hughes:2023apl}
M.R.R.~Hughes, S.D.~Mathur and M.~Mehta, \emph{{Lifting of two-mode states in
  the D1-D5 CFT}}, \href{https://doi.org/10.1007/JHEP01(2024)183}{\emph{JHEP}
  {\bfseries 01} (2024) 183}
  [\href{https://arxiv.org/abs/2309.03321}{{\ttfamily 2309.03321}}].

\bibitem{Hughes:2023fot}
M.R.R.~Hughes, S.D.~Mathur and M.~Mehta, \emph{{Lifting of superconformal
  descendants in the D1-D5 CFT}},
  \href{https://doi.org/10.1007/JHEP04(2024)129}{\emph{JHEP} {\bfseries 04}
  (2024) 129} [\href{https://arxiv.org/abs/2311.00052}{{\ttfamily
  2311.00052}}].

\bibitem{Gaberdiel:2023lco}
M.R.~Gaberdiel, R.~Gopakumar and B.~Nairz, \emph{{Beyond the Tensionless Limit:
  Integrability in the Symmetric Orbifold}},
  \href{https://arxiv.org/abs/2312.13288}{{\ttfamily 2312.13288}}.

\bibitem{Frolov:2023pjw}
S.~Frolov and A.~Sfondrini, \emph{{Comments on Integrability in the Symmetric
  Orbifold}},  \href{https://arxiv.org/abs/2312.14114}{{\ttfamily 2312.14114}}.

\bibitem{Mukund:2024}
C.~Ferko, S.~Murthy and M.~Rangamani, \emph{{Superstring partition functions in
  AdS$_3$}}, {\emph{To appear} (2024) }.

\bibitem{Gaberdiel:2021kkp}
M.R.~Gaberdiel, B.~Knighton and J.~Vo\v{s}mera, \emph{{D-branes in AdS$_{3}$
  \texttimes{} S$^{3}$ \texttimes{} \ensuremath{\mathbb{T}}$^{4}$ at k = 1 and
  their holographic duals}},
  \href{https://doi.org/10.1007/JHEP12(2021)149}{\emph{JHEP} {\bfseries 12}
  (2021) 149} [\href{https://arxiv.org/abs/2110.05509}{{\ttfamily
  2110.05509}}].

\bibitem{Maldacena:2001ss}
J.M.~Maldacena, G.W.~Moore and N.~Seiberg, \emph{{D-brane charges in five-brane
  backgrounds}},
  \href{https://doi.org/10.1088/1126-6708/2001/10/005}{\emph{JHEP} {\bfseries
  10} (2001) 005} [\href{https://arxiv.org/abs/hep-th/0108152}{{\ttfamily
  hep-th/0108152}}].

\bibitem{Polchinski:1998rr}
J.~Polchinski, \emph{{String theory. Vol. 2: Superstring theory and beyond}},
  Cambridge Monographs on Mathematical Physics, Cambridge University Press (12,
  2007),
  \href{https://doi.org/10.1017/CBO9780511618123}{10.1017/CBO9780511618123}.

\bibitem{Polchinski:1998rq}
J.~Polchinski, \emph{{String theory. Vol. 1: An introduction to the bosonic
  string}}, Cambridge Monographs on Mathematical Physics, Cambridge University
  Press (12, 2007),
  \href{https://doi.org/10.1017/CBO9780511816079}{10.1017/CBO9780511816079}.

\bibitem{Giveon:1994fu}
A.~Giveon, M.~Porrati and E.~Rabinovici, \emph{{Target space duality in string
  theory}}, \href{https://doi.org/10.1016/0370-1573(94)90070-1}{\emph{Phys.
  Rept.} {\bfseries 244} (1994) 77}
  [\href{https://arxiv.org/abs/hep-th/9401139}{{\ttfamily hep-th/9401139}}].

\bibitem{Borsato:2023dis}
R.~Borsato, \emph{{Lecture notes on current-current deformations}},
  \href{https://arxiv.org/abs/2312.13847}{{\ttfamily 2312.13847}}.

\end{thebibliography}\endgroup
\end{document}